\documentclass{aa}
\usepackage{graphicx}
\usepackage{natbib}
\usepackage{txfonts}

\begin{document}

\title{The first study of four doubly eclipsing systems}

\author{Zasche,~P.~\inst{1},
        Henzl,~Z.~\inst{2},
        K\'ara,~J.~\inst{1}}

\offprints{Petr Zasche, \email{zasche@sirrah.troja.mff.cuni.cz}}

 \institute{
  $^{1}$ Charles University, Faculty of Mathematics and Physics, Astronomical Institute, V~Hole\v{s}ovi\v{c}k\'ach 2, CZ-180~00, Praha 8, Czech Republic\\
  $^{2}$ Variable Star and Exoplanet Section of the Czech Astronomical Society, Vset\'{\i}nsk\'a 941/78, CZ-757 01 Vala\v{s}sk\'e Mezi\v{r}\'{\i}\v{c}\'{\i}, Czech Republic
 }

\titlerunning{Study of doubly eclipsing systems}
\authorrunning{Zasche et al.}

  \date{Received \today; accepted ???}

\abstract{We present the discovery and the very first analysis of four stellar systems showing two
periods of eclipses, that are the objects classified as doubly eclipsing systems. Some of them were
proved to orbit each other thanks to their eclipse-timing-variations (ETVs) of both pairs, hence they
really constitute rare quadruples with two eclipsing pairs. Some of them do not, as we are still
waiting for more data to detect their mutual movement. Their light curves and period changes were
analysed. All of them are detached and near-contact, but none of them contact; moreover, to our
knowledge none of these stars can be considered as blend of two spatially unresolved close components
on the sky. These systems are CzeV2647 (0.5723296 + 0.9637074 days), proved to orbit with 4.5-year
periodicity; CzeV1645 (1.0944877 + 1.6594641 days), with a rather questionable detection of ETV;
CzeV3436 (0.6836870 + 0.3833930 days); and, finally, OGLE SMC-ECL-1758 (0.9291925 + 3.7350826 days),
proved to move on its 30-year orbit. Even more surprising is the fact that most of these systems show
the ratio of their two orbital periods close to coupling near some resonant values of small integers,
namely CzeV2647, with only 1\% from 3:5 resonance, CzeV1645 1\% from 2:3 resonance, and OGLE
SMC-ECL-1758 with only 0.49\% from 1:4 resonance. The nature of these near-resonant states still
remains a mystery.}


\maketitle

\section{Introduction} \label{intro}

The role of classical eclipsing binaries in current astrophysical research has been discussed elsewhere
(see for example \citealt{2012ocpd.conf...51S}). However, in the era of huge surveys and extremely
precise satellite data, the objects with double eclipsing features can also be discovered and studied.
The use of such objects brings us new insight into stellar formation mechanisms (see for instance
\citealt{2021Univ....7..352T}). However, many more constraints should be taken into account when
dealing with two binaries in one system - independent analysis should lead to the same distance, same
age, same metallicity, and so on.

The doubly eclipsing binaries as an independent topic is still rather new. The first of its kind
(V994~Her) was discovered by \cite{2008MNRAS.389.1630L}. It was about ten years later when the mutual
orbit of the two pairs was definitively confirmed; hence, it comprised a still rather rare 2+2
quadruple system. This is quite a typical situation due to the long time taken to collect the data
needed for the confirmation of the mutual orbit via eclipse-timing-variation (ETV). The only
comprehensive study of such fascinating objects is that published by our group in
\citet{2019A&A...630A.128Z}. Since then, several new discoveries of doubly eclipsing objects have been
published in the last years, mostly based on new and extremely precise data from the TESS satellite
\citep{2015JATIS...1a4003R}.

The complex nature of these quadruple systems make them ideal astrophysical laboratories, especially for
the celestial mechanics and the dynamical evolution of their orbits, the role of their orbit-orbit interaction, the
role of Kozai cycles, the long-term evolution of the orbits such as the precession of the two eclipsing pairs,
or the detection of small changes of other their orbital parameters. All of these are ongoing
problems and only a thorough dynamical modelling would we be able to answer all these questions. However,
for a proper modelling, one needs precise input data concerning individual physical parameters of the
components, as well as of their orbits. Hence, our present study should be considered as a
starting point in such an analysis.

\section{System selection}

Most of the systems showing two sets of eclipses published in the last few years were detected in the
TESS data (and sometimes also proved to orbit each other after carefully analysing the photometric TESS
data). These data have a huge advantage not only due to their precision, but also thanks to their
cadency and time coverage. For these reasons, the detected periods of their mutual orbits are usually
rather short (sometimes only hundreds of days, see \citealt{2021ApJ...917...93K}), but the depths of
eclipses are sometimes rather shallow. What is obviously not a problem for TESS photometry can
sometimes be quite tricky for a normal ground-based observatories.

Knowing all these reasons and limitations of the TESS data (only a short time span is currently
covered), we decided to follow a different method. Our new discovered systems on the northern sky are
those found by chance as a by-product in some fields already monitored with our small-aperture amateur
telescopes located in the Czech Republic and controlled by one of the authors (Z.H.). The one
complementary system from the southern sky is located in the SMC galaxy and was included in this study
due to the fact that our photometric monitoring of the target during the last years finally led to
definite proof that both the ETVs show evidence of their orbit, hence constituting a real quadruple.

Some of the candidates for new doubly eclipsing systems were omitted due to a possibility that the two
periodic signals come from different sources on the sky. When two very close companions are located on
the sky, and we are not able to define the common origin with our technique, we skip such a star
for the moment. For all of the presented systems, no close companion is evident in various sky surveys,
and such systems are very likely doubly eclipsing quadruples.

\begin{table*}
  \caption{Information about systems under analysis.}  \label{systemsInfo}
  \scalebox{0.94}{
  \begin{tabular}{c c c c c c}\\[-6mm]
\hline \hline
  Target name         &  Other name             &RA [J2000.0]& DE [J2000.0]& Mag$_{max}$ $^*$ &  Temperature information      \\
 \hline
 CzeV2647             & UCAC4 642-089788        & 20 14 05.6 & +38 22 58.5 &  13.43 (V)       & T$_{eff} =  6404$ K \citep{2019AJ....158...93B} \\
 CzeV1645             & UCAC4 755-079509        & 23 23 32.1 & +60 53 13.0 &  13.75 (V)       & T$_{eff} =  5400$ K \citep{2018AaA...616A...1G} \\
 CzeV3436             & UCAC4 631-070565        & 19 21 24.9 & +36 07 58.9 &  14.72 (V)       & T$_{eff} =  6817$ K \citep{2019AaA...628A..94A} \\
 OGLE SMC-ECL-1758    & MACHO 212.15962.70      & 00 49 55.2 & -73 16 51.3 &  16.19 (I)       & T$_{eff} = 11162$ K \citep{2019AaA...628A..94A} \\
 \hline
\end{tabular}}\\
 {\small Notes: $^*$ - Out-of-eclipse magnitude, $V_{mag}$ from UCAC4 catalogue \citep{2013AJ....145...44Z}, or Guide Star Catalog II \citep{2008AJ....136..735L}, $I_{mag}$ from OGLE survey \citep{2013AcA....63..323P}.}
\end{table*}

\section{Data availability}

Dealing only with the photometry, we mainly used the TESS data, together with our own ground-based
ones. The TESS photometry was extracted using the raw TESS data with the {\tt{lightkurve}}
tool \citep{2018ascl.soft12013L}. The usual problem with too large TESS pixels (hence a possible influence
of other close-by targets in the vicinity of the main star), other than the necessity of additional
flattening of the TESS fluxes, which is discribed elsewhere \citep{2019PASP..131i4502F}, one has to be
cautious with the interpretation of the results and its comparison with other photometric data.

The ground-based observations were obtained by one of the authors, Z.H., at his private observatory in
Velt\v{e}\v{z}e u Loun, Czech Republic. A quite untypical observational setup was used due to the filter
used for that data, namely the astrophotographic Baader UV/IR Sperrfilter. It very effectively cuts
almost all signal below 400nm and above 700nm. Maybe the most similar among other standard photometric
filters is the transmission curve of the former SuperWASP filter \citep{2006PASP..118.1407P}.
Therefore, for the proper modelling of the limb-darkening in the {\sc PHOEBE} light-curve analysis
program \citep{2005ApJ...628..426P}, we adopted coefficients for the SWASP band in our modelling, which
we do not expect to affect our analysis significantly.

For the one southern object, the follow-up data were secured using the 1.54-m Danish telescope located
in La Silla, Chile. A standard $I$ band Cousins filter was used \citep{1990PASP..102.1181B}. For the
reduction of the data, the dark frames, flat fields, and standard aperture photometry were used to
extract the data.

\section{Analysis}

Due to the fact that our whole analysis is solely based on photometry (unfortunately, the systems are
rather faint for obtaining good spectra), we used the program {\sc PHOEBE}, which is based on the
Wilson-Devinney method \citep{1971ApJ...166..605W}. It uses classical Roche geometry, and this approach
was adequate for our study. However, several simplifying assumptions have to be made due to missing
spectroscopy. The problem of deriving the mass ratio for detached binaries solely from the photometry
was discussed elsewhere (see for instance \cite{2005Ap&SS.296..221T}). However, for the precise TESS
photometry and more compact binaries with large out-of-eclipse ellipsoidal variations, we can also try
to fit the mass ratio as a free parameter.

Another way to simplify assumptions is to keep the synchronous rotation (that is F=1.0), albedo, and
gravity brightening fixed at their suggested values according to a particular temperature. The second
most important issue is the temperature. Due to not having known the spectral types of either of the
components, we used an effective temperature estimation based on GAIA data \citep{2016A&A...595A...1G}
(see our Table \ref{systemsInfo}). We are aware of the fact that these numbers could be shifted from
reality due to several effects such as interstellar extinction or a combination of signals from four
components into one flux, which we then analysed with a simple single-component approach. However, this
rough estimation is still better than using a reddened $(B-V)$ index, and the resulting temperature
ratio should at least be viable in this aspect.

Our entire approach comes from the fact that both the periods of these doubly eclipsing systems are
already known (from our rough photometric data), and we can plot the phased light curves with both
periods. Our method was the following. The dominant eclipsing binary (pair A) was first modelled in
{\sc PHOEBE}, but still using complete photometry of both pairs together. Then, this model was
subtracted from the original photometry. On the residuals, pair B was analysed and modelled. With this
model, we made a preliminary period analysis using our automatic-fitting-procedure (AFP) method
\citep{2014A&A...572A..71Z}. This approach uses a light-curve template on shorter time intervals to
derive both primary and secondary times of eclipses simply via shifting the template over the data in
the x and y axes. With the resulting times of eclipses, one can easily see whether curvature of the ETV
curve is evident or not. One can obviously also correct the orbital period of pair B. With this
corrected period, a new analysis of pair B was carried out. Such a fit of pair B was then subtracted
from the complete photometry of both pairs. Hence, we obtain only the residuals containing solely pair
A data. The analysis of pair A was done in a similar way. This approach was repeated several times to
obtain really symmetric and flat residuals when subtracting both pairs from the complete photometry.
Obviously, when the curvature of the ETV is too large and adequately short, we have to use  different
ephemerides and orbital periods for  different epochs of data, and the whole process is repeated
independently several times on shorter time intervals.

The fitting process of light-curve analysis starting from the easiest assumption that both pairs are
similar, that is third light is 50\% for both. Then, this assumption was left and the third light was
also set free and computed as an independent parameter. However, its significance is rather low in poor
data and these ones with only shallow eclipses. Naturally, the resulting third light ratio should be in
agreement with the resulting mass ratio of both pairs from our analysis. However, this is sometimes
rather problematic due to additional light in the aperture in large TESS pixels.

\section{Results}

Here, we focus on the individual systems of our analysis in more detail.

\subsection{CzeV2647}

The very first system in our analysis is named CzeV2647, or UCAC4 642-089788, discovered as a variable
about a year ago by the author Z.H. at his private observatory. The dominant variation with a period of
about 0.57 days is also clearly visible on the data from surveys ASAS-SN (\citealt{2014ApJ...788...48S},
and \citealt{2017PASP..129j4502K}), ZTF \citep{2019PASP..131a8003M}, and of course by TESS. The object
was discovered by chance in some already monitored fields and was only recognised as a variable
from the ATLAS survey \citep{2018AJ....156..241H}, which gave an incorrect period about half of the true
one. Due to its low brightness, unfortunately only very little is known about the star. Even the
temperature estimations range between 4900 and 8530 K. Its distance seems to be about 1 kpc from the
Sun \citep{2021AJ....161..147B}.

\begin{table*}
\caption{Derived parameters for the two inner binaries A and B.}
 \label{TabLC}
 \tiny
  \centering \scalebox{0.999}{
\begin{tabular}{c |c c|c c| c c| c}
   \hline\hline\noalign{\smallskip}
 \multicolumn{1}{c|}{System   } & \multicolumn{2}{c|}{CzeV2647}          & \multicolumn{2}{c|}{CzeV1645}         & \multicolumn{2}{c|}{CzeV3436}          & OGLE SMC-ECL-1758 \\ \hline
  \multicolumn{1}{c|}{ }        &  \multicolumn{7}{c}{{\sc p a i r \,\, A}} \\ \hline
 \multicolumn{1}{c|}{Light curve} &  TESS           &      Z.Henzl       &   TESS            &   Z.Henzl         &    TESS            &   Z.Henzl         &  OGLE IV          \\ \hline
  $i$ [deg]                     & 78.63 $\pm$ 0.16  &  87.81 $\pm$ 0.37  &  83.29 $\pm$ 0.33 & 84.42 $\pm$ 0.31  &  63.92 $\pm$ 0.47  & 61.78 $\pm$ 0.85  & 72.12  $\pm$ 0.73 \\
 $q=\frac{M_2}{M_1}$            & 0.257 $\pm$ 0.007 &  0.415 $\pm$ 0.018 & 0.949 $\pm$ 0.025 & 0.998 $\pm$ 0.021 & 0.940 $\pm$ 0.026  & 0.898 $\pm$ 0.037 &  0.63 $\pm$ 0.05  \\
 $T_1$ [K]                      &  6404 (fixed)     &    6404 (fixed)    &  5400 (fixed)     &   5400 (fixed)    &    6817 (fixed)    &    6817 (fixed)   &  11162 (fixed)    \\
 $T_2$ [K]                      &  4459 $\pm$ 124   &   4639 $\pm$ 257   & 4998 $\pm$ 112    & 4956 $\pm$ 65     & 4546 $\pm$ 159     &  4503 $\pm$ 201   & 8457 $\pm$ 258    \\
 $R_1/a$                        & 0.449 $\pm$ 0.002 &  0.420 $\pm$ 0.009 & 0.314 $\pm$ 0.003 & 0.315 $\pm$ 0.005 & 0.372 $\pm$ 0.005  & 0.384 $\pm$ 0.007 & 0.368 $\pm$ 0.005 \\
 $R_2/a$                        & 0.199 $\pm$ 0.001 &  0.247 $\pm$ 0.007 & 0.260 $\pm$ 0.003 & 0.225 $\pm$ 0.004 & 0.351 $\pm$ 0.008  & 0.333 $\pm$ 0.004 & 0.286 $\pm$ 0.009 \\
 $L_1$ [\%]                     & 60.1 $\pm$ 0.6    &   42.9 $\pm$ 3.4   &  29.0 $\pm$ 0.6   & 39.4 $\pm$ 1.3    & 38.0 $\pm$ 0.6     & 38.6 $\pm$ 1.0    & 40.1 $\pm$ 3.2    \\
 $L_2$ [\%]                     &  3.5 $\pm$ 0.5    &    2.9 $\pm$ 0.9   &  16.7 $\pm$ 0.8   & 11.6 $\pm$ 1.1    &  8.5 $\pm$ 0.5     &  3.6 $\pm$ 0.4    & 15.0 $\pm$ 1.0    \\
 $L_3$ [\%]                     & 36.4 $\pm$ 1.2    &   54.2 $\pm$ 7.2   &  54.3 $\pm$ 2.4   & 49.0 $\pm$ 2.2    & 53.5 $\pm$ 1.4     & 57.8 $\pm$ 1.9    & 44.9 $\pm$ 3.8    \\ \hline
  \multicolumn{1}{c|}{ }     & \multicolumn{7}{c}{{\sc p a i r \,\, B}}\\ \hline
  \multicolumn{1}{c|}{Light curve} &   TESS         &      Z.Henzl       &    TESS           &    Z.Henzl        &     TESS           &   Z.Henzl         &  OGLE IV          \\ \hline
 $i$ [deg]                      &  68.04 $\pm$ 0.34 &  76.71 $\pm$ 1.59  &  83.20 $\pm$ 0.81 & 84.61 $\pm$ 0.26  & 60.41 $\pm$ 0.52   & 57.80 $\pm$ 0.64  & 88.54 $\pm$ 1.20  \\
 $q=\frac{M_2}{M_1}$            & 1.011 $\pm$ 0.029 &  0.835 $\pm$ 0.047 & 0.933 $\pm$ 0.043 & 0.970 $\pm$ 0.025 & 1.033 $\pm$ 0.039  & 1.020 $\pm$ 0.042 &  0.62 $\pm$ 0.07  \\
 $T_1$ [K]                      &  6404 (fixed)     &   6404 (fixed)     & 5400 (fixed)      &  5400 (fixed)     &   6817 (fixed)     &    6817 (fixed)   &   11162 (fixed)   \\
 $T_2$ [K]                      &  6379 $\pm$ 238   & 6218 $\pm$ 345     &  5111 $\pm$ 103   & 5096 $\pm$ 60     & 6224 $\pm$ 235     & 6135 $\pm$ 277    & 10675 $\pm$ 241   \\
 $R_1/a$                        & 0.274 $\pm$ 0.006 & 0.287 $\pm$ 0.010  & 0.223 $\pm$ 0.007 & 0.223 $\pm$ 0.002 & 0.380 $\pm$ 0.008  & 0.375 $\pm$ 0.009 & 0.337 $\pm$ 0.005 \\
 $R_2/a$                        & 0.186 $\pm$ 0.007 & 0.093 $\pm$ 0.009  & 0.209 $\pm$ 0.005 & 0.213 $\pm$ 0.003 & 0.385 $\pm$ 0.007  & 0.383 $\pm$ 0.009 & 0.155 $\pm$ 0.008 \\
 $L_1$ [\%]                     & 25.9  $\pm$ 1.4   & 20.6 $\pm$ 1.8     & 28.7 $\pm$ 0.3    & 30.6 $\pm$ 0.7    & 30.0 $\pm$ 1.0     & 25.2 $\pm$ 1.3    & 39.0 $\pm$ 2.0    \\
 $L_2$ [\%]                     & 11.7  $\pm$ 3.2   & 1.9 $\pm$ 1.0      & 18.3 $\pm$ 1.0    & 18.6 $\pm$ 0.5    & 23.8 $\pm$ 0.9     & 18.0 $\pm$ 1.2    &  6.8 $\pm$ 0.8    \\
 $L_3$ [\%]                     & 62.4  $\pm$ 1.9   & 77.5 $\pm$ 2.3     & 53.0 $\pm$ 1.7    & 50.9 $\pm$ 1.4    & 46.2 $\pm$ 2.3     & 56.8 $\pm$ 2.4    & 54.2 $\pm$ 3.6    \\
 \noalign{\smallskip}\hline
\end{tabular}} \\
  \begin{flushleft}
  \footnotesize Note: Individual uncertainties of parameters taken from {\sc PHOEBE} only, which are usually underestimated.\\
  \end{flushleft}
\end{table*}

\begin{table*}
\caption{Results of ETV analysis for our studied systems.}
 \label{TabETV}
 \tiny
  \centering \scalebox{0.999}{
\begin{tabular}{c | c c c c}
   \hline\hline\noalign{\smallskip}
 \multicolumn{1}{c|}{    }  &        CzeV2647           &         CzeV1645          &         CzeV3436          & OGLE SMC-ECL-1758         \\ \hline
  $JD_{0,A}$ [HJD-2450000]  & 7853.0554 $\pm$ 0.0008    & 8785.1259 $\pm$ 0.0006    & 8699.0803 $\pm$ 0.0010    & 5000.4532 $\pm$ 0.0037    \\
 $P_A$ [day]                & 0.5723295 $\pm$ 0.0000005 & 1.0944876 $\pm$ 0.0000016 & 0.6836872 $\pm$ 0.0000013 & 0.9291937 $\pm$ 0.0000011 \\
  $JD_{0,B}$ [HJD-2450000]  & 7853.5475 $\pm$ 0.0026    & 8784.6148 $\pm$ 0.0004    & 8699.8441 $\pm$ 0.0005    & 5003.0698 $\pm$ 0.0047    \\
 $P_B$ [day]                & 0.9637074 $\pm$ 0.0000037 & 1.6594636 $\pm$ 0.0000015 & 0.3833927 $\pm$ 0.0000004 & 3.7350755 $\pm$ 0.0000058 \\
 $p_{AB}$ [yr]              &  4.55 $\pm$ 0.65          &           --              &          --               & 30.3 $\pm$ 1.2            \\
 $A_A$ [d]                  & 0.0093 $\pm$ 0.0005       &           --              &          --               & 0.0672 $\pm$ 0.0019       \\
 $A_B$ [d]                  & 0.0140 $\pm$ 0.0027       &           --              &          --               & 0.0641 $\pm$ 0.0037       \\
 $e$                        & 0.602 $\pm$ 0.019         &           --              &          --               & 0.642 $\pm$ 0.004         \\
 $\omega$ [deg]             & 29.3 $\pm$ 20.8           &           --              &          --               & 61.9 $\pm$ 3.7            \\
 $T_0$ [HJD-2450000]        & 5642 $\pm$ 451            &           --              &          --               & 2219 $\pm$ 54             \\
 \noalign{\smallskip}\hline
\end{tabular}} \\
\end{table*}

\begin{figure}
 \centering
 \begin{picture}(280,150)
 \put(0,80){
  \includegraphics[width=0.24\textwidth]{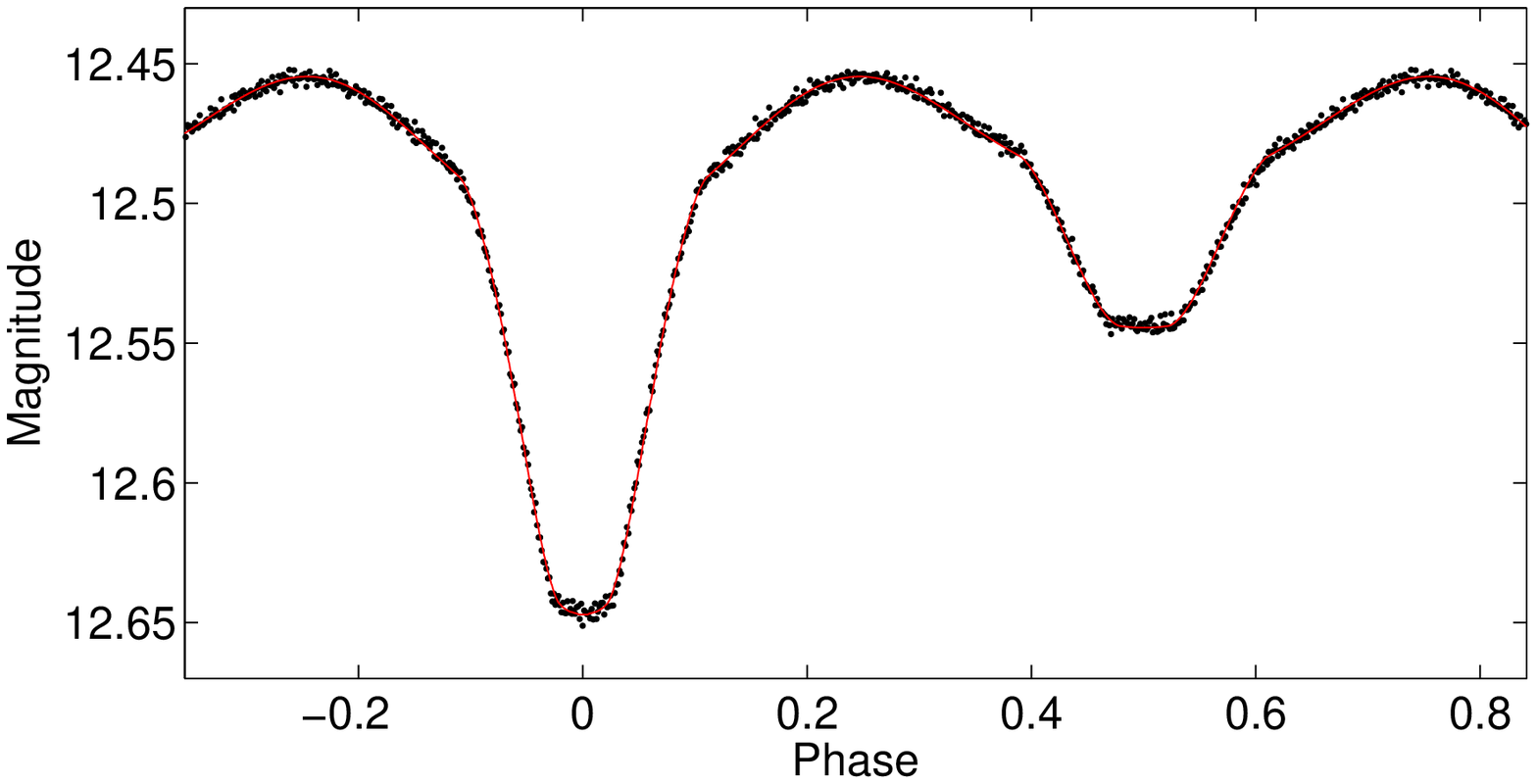}}
  \put(0,0){
  \includegraphics[width=0.24\textwidth]{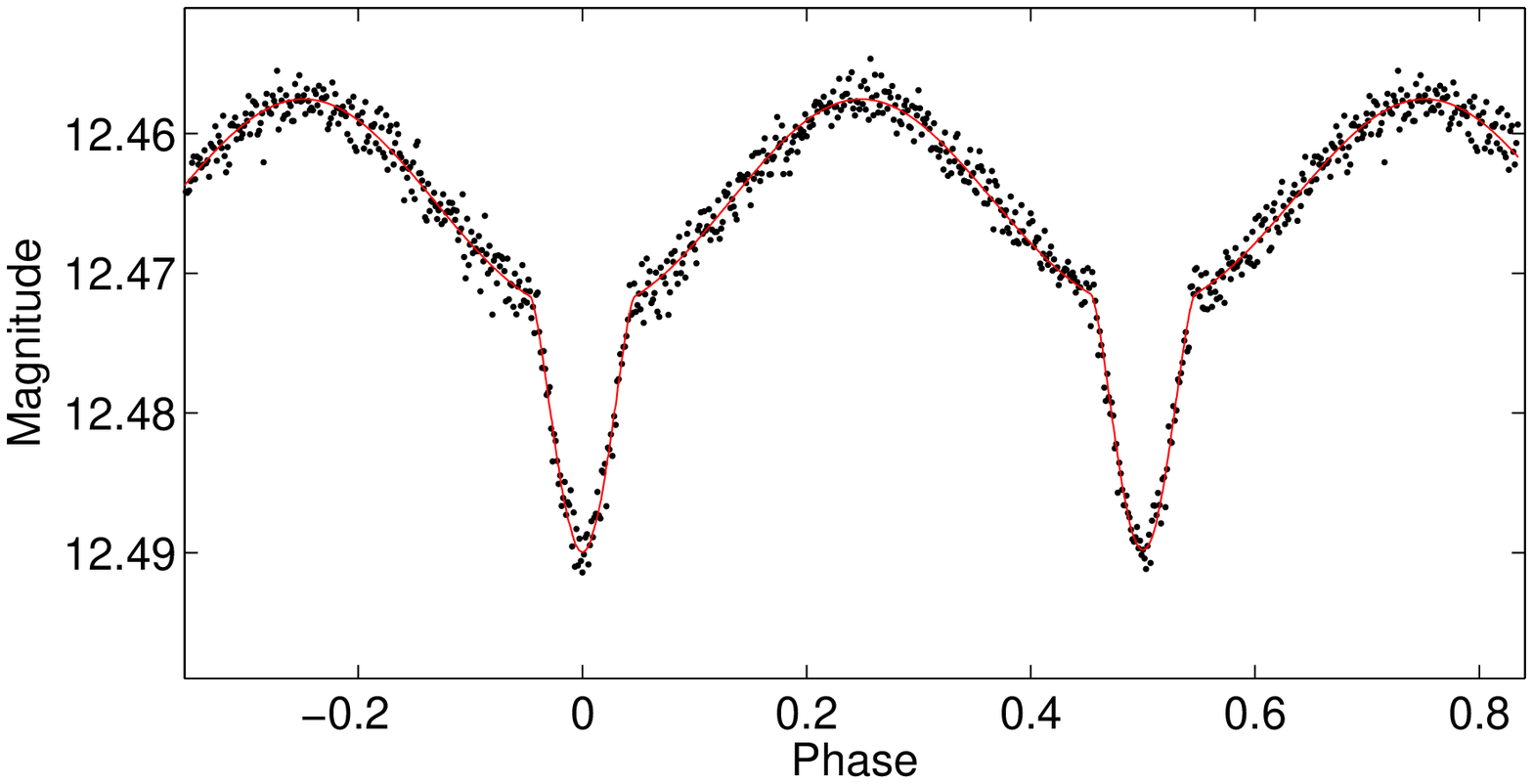}}
  \put(128,80){
  \includegraphics[width=0.24\textwidth]{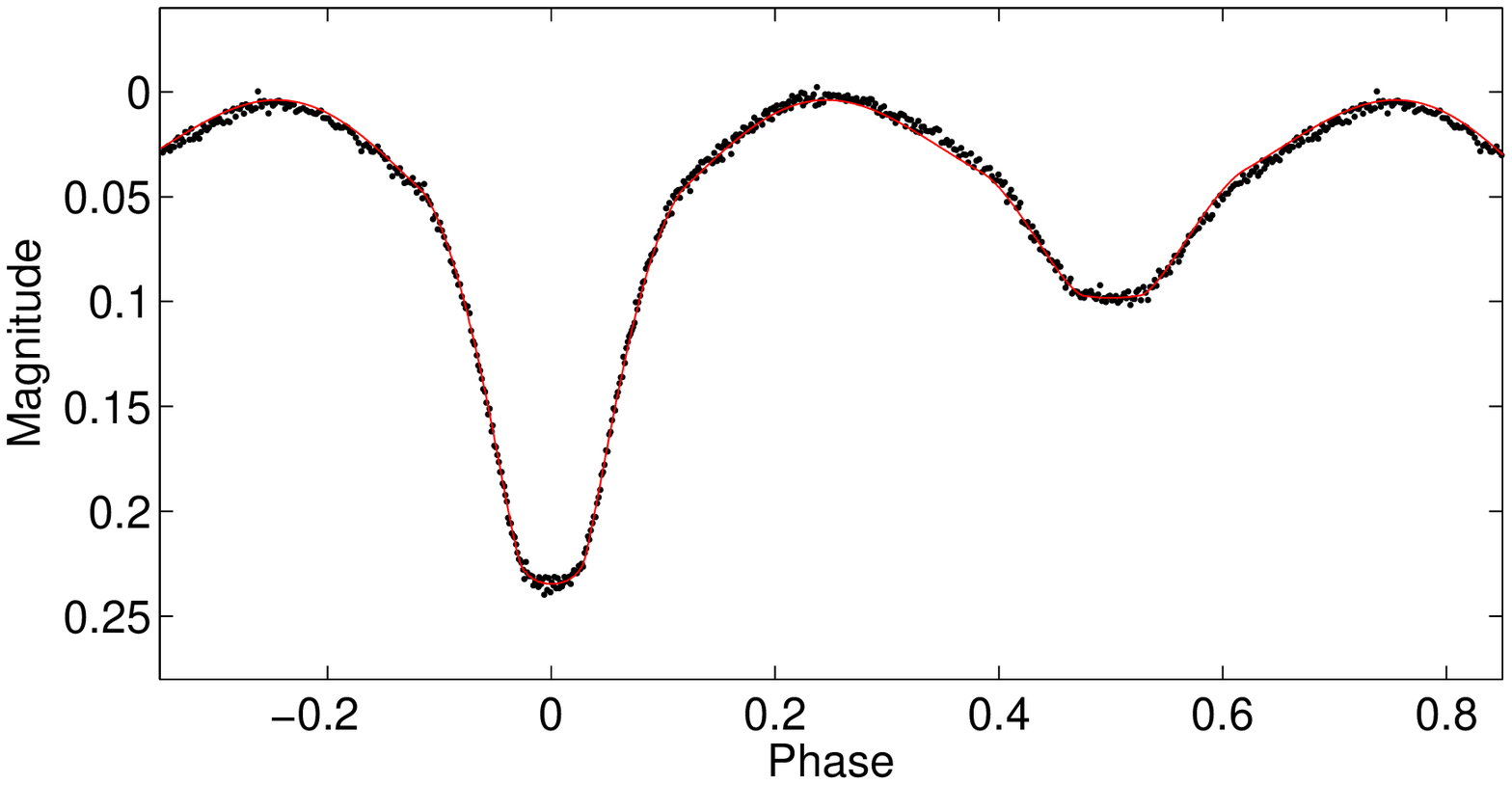}}
  \put(128,0){
  \includegraphics[width=0.24\textwidth]{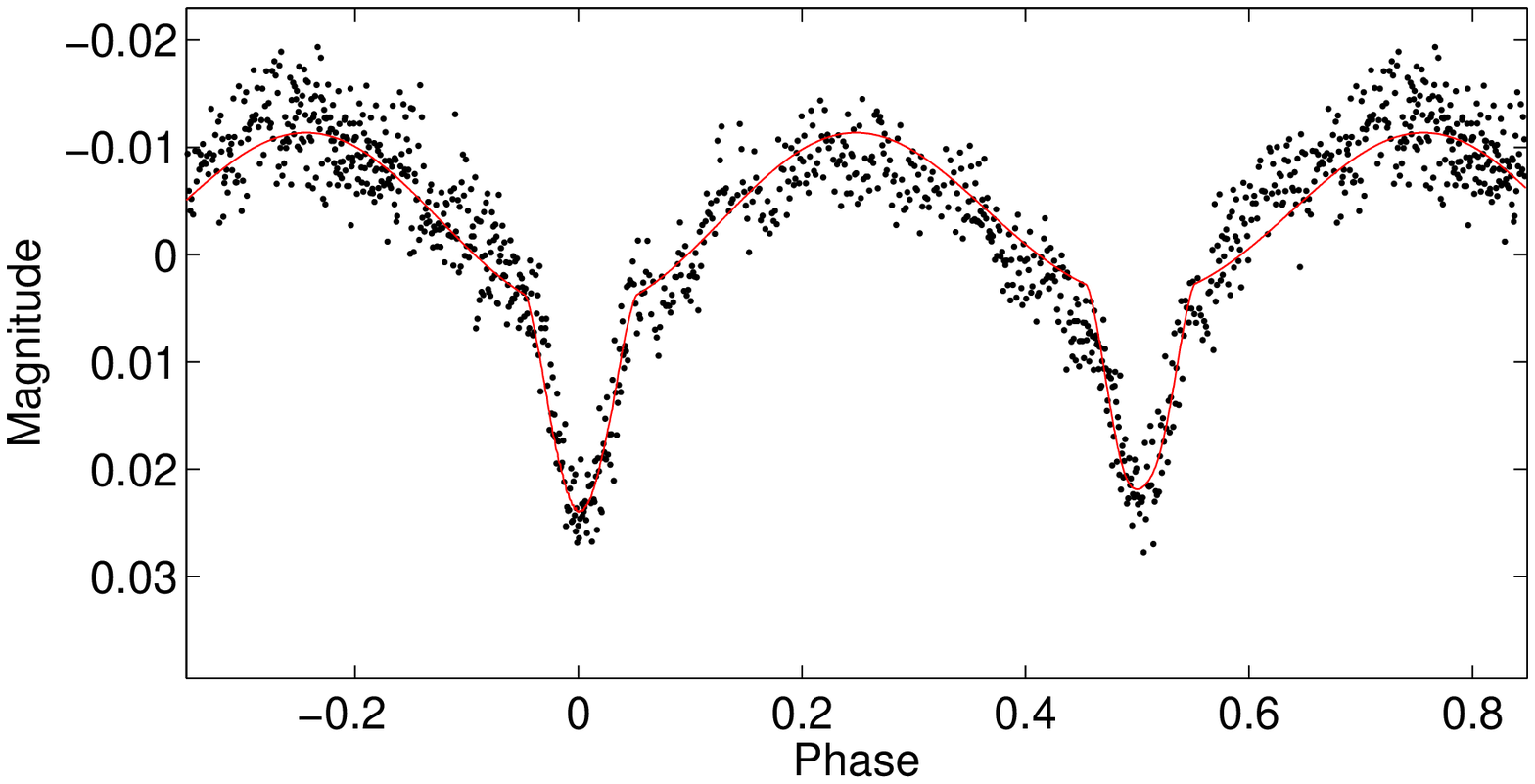}}
   \put(65,95){\small TESS: Pair A}
   \put(56,12){\small TESS: Pair B}
   \put(195,95){\small Z.H.: Pair A}
   \put(190,12){\small Z.H.: Pair B}
  \end{picture}
  \caption{Light-curve fits of CzeV2647 of both eclipsing pairs, as resulting from PHOEBE (see the text for details). TESS data are compared with our ground-based data  obtained by one of the authors (Z.H.).}
  \label{FigLC_czev2647}
 \end{figure}

We collected all available data for the star, finding that the best for the analysis are our own
ground-based data together with the TESS satellite data (especially Sector 41 with 2-min cadence).
The latter is for the precise light-curve analysis to derive the physical parameters, while the
former was used to derive many useful times of eclipses of both pairs for ETV analysis, proving that the
system is a real quadruple one. As one can see from the final light-curve fits (see Fig.
\ref{FigLC_czev2647}), both the binaries show very different eclipse depths. Pair A shows about five times deeper eclipses, which is the reason why the star was being missed as a doubly eclipsing one.

\begin{figure}
 \centering
 \includegraphics[width=0.40\textwidth]{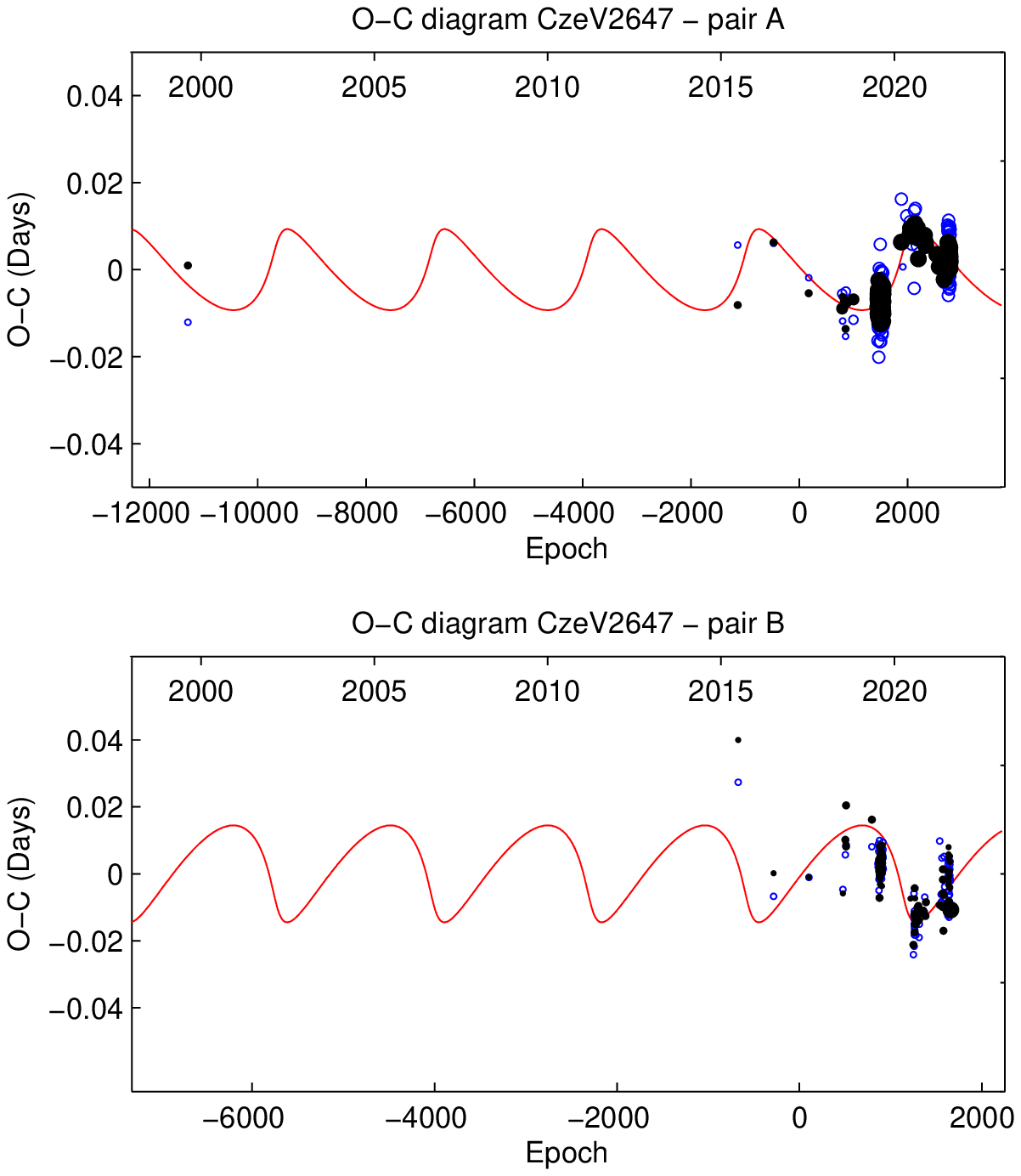}
 \caption{ETV diagram of CzeV2647 showing two sets of eclipses of A and B pairs as resulting from our analysis. Theoretical curve representing the fit given in Table \ref{TabETV} is plotted as a red solid curve, while the black dots stand for primary eclipses and the open circles for secondary ones: the larger the symbol, the higher the weight.}
 \label{Fig_OC_czev2647}
\end{figure}

The analysis in {\sc PHOEBE} yielded the parameters given in Table \ref{TabLC}, showing that we are dealing with two detached binaries. Pair A is the dominating one in luminosity (from photometry) as
well as in mass (from ETV, see below). Due to good coverage and precision of the TESS data for pair A
and its deep eclipses, we also tried to fit the mass ratio here. As can be seen from Table \ref{TabLC},
such a task also for the shallow pair B is much more tricky, and the result is affected by large errors.
Owing to the missing spectroscopy, only relative values of radii can be derived properly. Moreover, the
star was found to be located close to the resonant ratio of its orbital periods at only 1\% from
3:5.

What is quite surprising is the fact that both solutions provided with TESS and our ground-based data
(see Fig. \ref{FigLC_czev2647} for comparison) resulted in quite different parameters, and the picture
of the system is still questionable. We are aware of the fact that the {\sc PHOEBE} code itself usually
provides rather unreliably small errors of individual parameters, but a real uncertainty of the
particular one can better be estimated as a difference between these two solutions. As one can see, values such as mass ratios or inclinations resulted in very different values, and hence the
solution should be taken with caution. However, its overall shape as detached binaries with the pair A
being the dominant one is in agreement within both these solutions. We can only speculate about the
origin of such a discrepancy. The ground-based data are of much worse quality; moreover, they also show slight asymmetry of both light curves outside of minima, which can only be fitted with a spot on
the surface. However, due to the non-detection of anything similar in the TESS data, we did not include
any such hypothesis in our analysis. Quite surprising is the fact that we detected much higher
contributing third light for both binaries from our ground-based photometry, despite the fact that it uses smaller
pixels and higher angular resolution than the TESS data. High values of both third lights are certainly
responsible for both inclination angles being higher than from TESS photometry due to their strong
correlation. However, we were not able to trace why is it so. The explanation may be that the
ground-based photometry itself is not precise enough for independent derivation of its mass ratio, and this value should also be kept fixed from the TESS solution. Fixing it to its proper value would also
shift the whole solution and maybe diminish the third-light inclination problem.

The long-term evolution of both A and B orbits can be studied using the eclipse timing data (so-called
ETV analysis, or the $O-C$ diagram analysis). We used the same approach in our former analysis of
doubly eclipsing systems CzeV1731 \citep{2020A&A...642A..63Z}, TIC 168789840
\citep{2021AJ....161..162P}, or all systems in \cite{2019A&A...630A.128Z}. Such a method is very
powerful and effective when good data are available, adequately distributed in time, and well covered for both
A and B binaries. This is partly true here for the dominant pair A, but quite problematic for pair B.
As one can see from Fig. \ref{Fig_OC_czev2647}, the variation of eclipses with the orbital period
of about 4.55-yr is definitely present in pair A. For pair B, this is still barely visible. The
very first eclipses around the year 2000 show us that no mass transfer is present in the system and its
orbital period is stable over longer time intervals. The only definitely detectable shapes of the ETV
for pair B are both periastron passages on the long orbit, but the overall shape of the variation is derived mainly from the precise pair-A data. The amplitudes of both ETVs of about 0.01 days are
about the minimum of what can be detected with our method with such a quality of data. The lower amplitude
of ETV for pair A is to be expected due to its higher luminosity (and hence also expected higher
mass), but the amplitude of pair B is still affected by large uncertainty now. More precise data
secured in the upcoming years would be very welcome to constrain the orbit with higher fidelity.

\subsection{CzeV1645}

The next stellar system discovered as suspected for doubly eclipsing nature is named CzeV1645, or UCAC4
755-079509. In the ASAS-SN database, it is also detected as variable and named ASASSN-V J232332.11+605312.9.
Hence, its dominant 1.09-day variation was already being recognized. We call this pair A, while another
periodicity with period of about 1.66 days was also detected. This is quite a typical situation - the
pair already detected is usually the one with larger photometric amplitude, or the one with shorter
orbital period. Only very limited information was published for this star, and it even lacked a proper
parallax value from the GAIA DR2. However, the TESS Input Catalog \citep{2019AJ....158..138S} gave its
distance as 1413 pc.

\begin{figure}
 \centering
 \begin{picture}(280,150)
 \put(0,80){
 \includegraphics[width=0.24\textwidth]{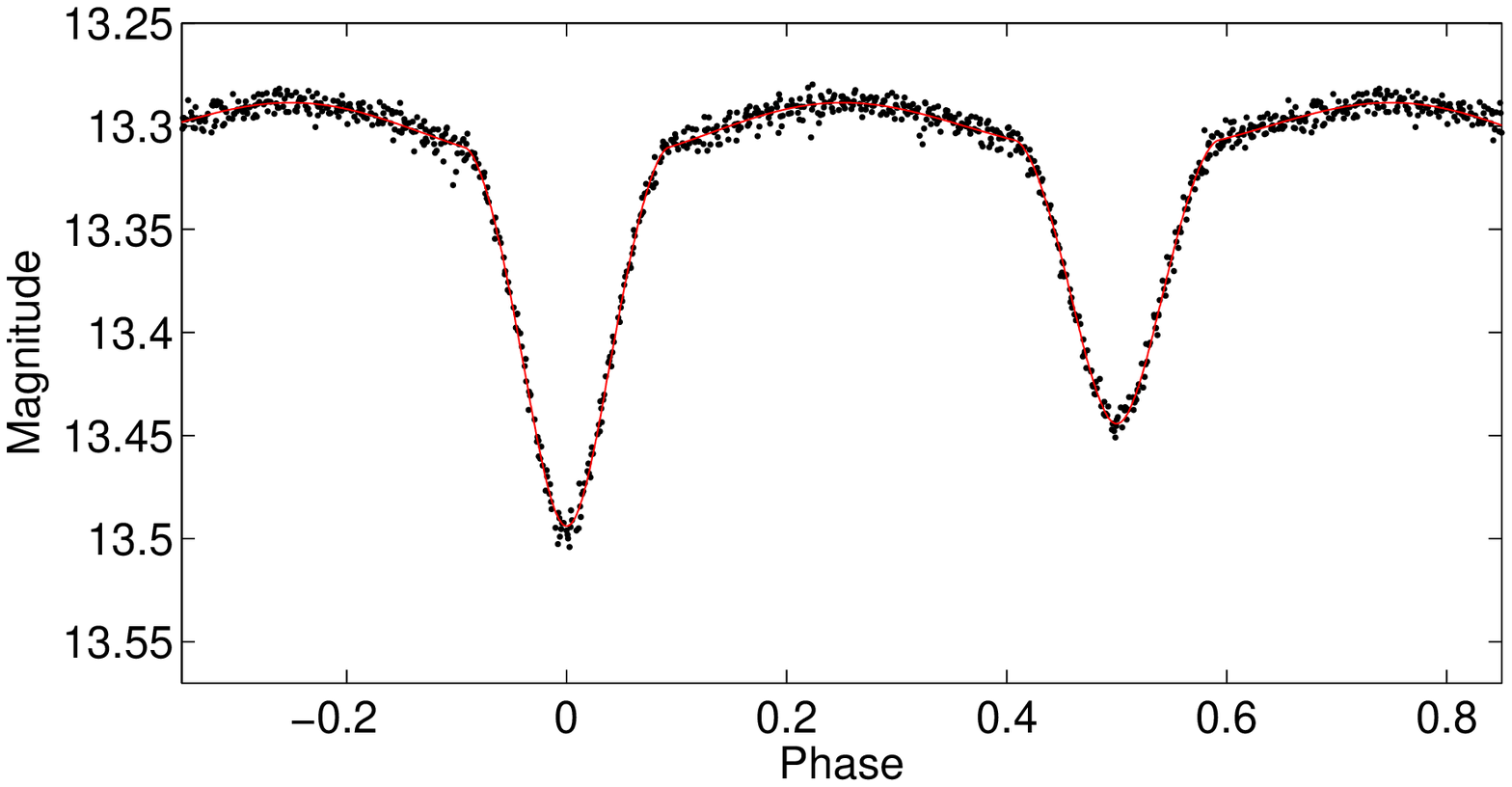}}
 \put(0,0){
 \includegraphics[width=0.24\textwidth]{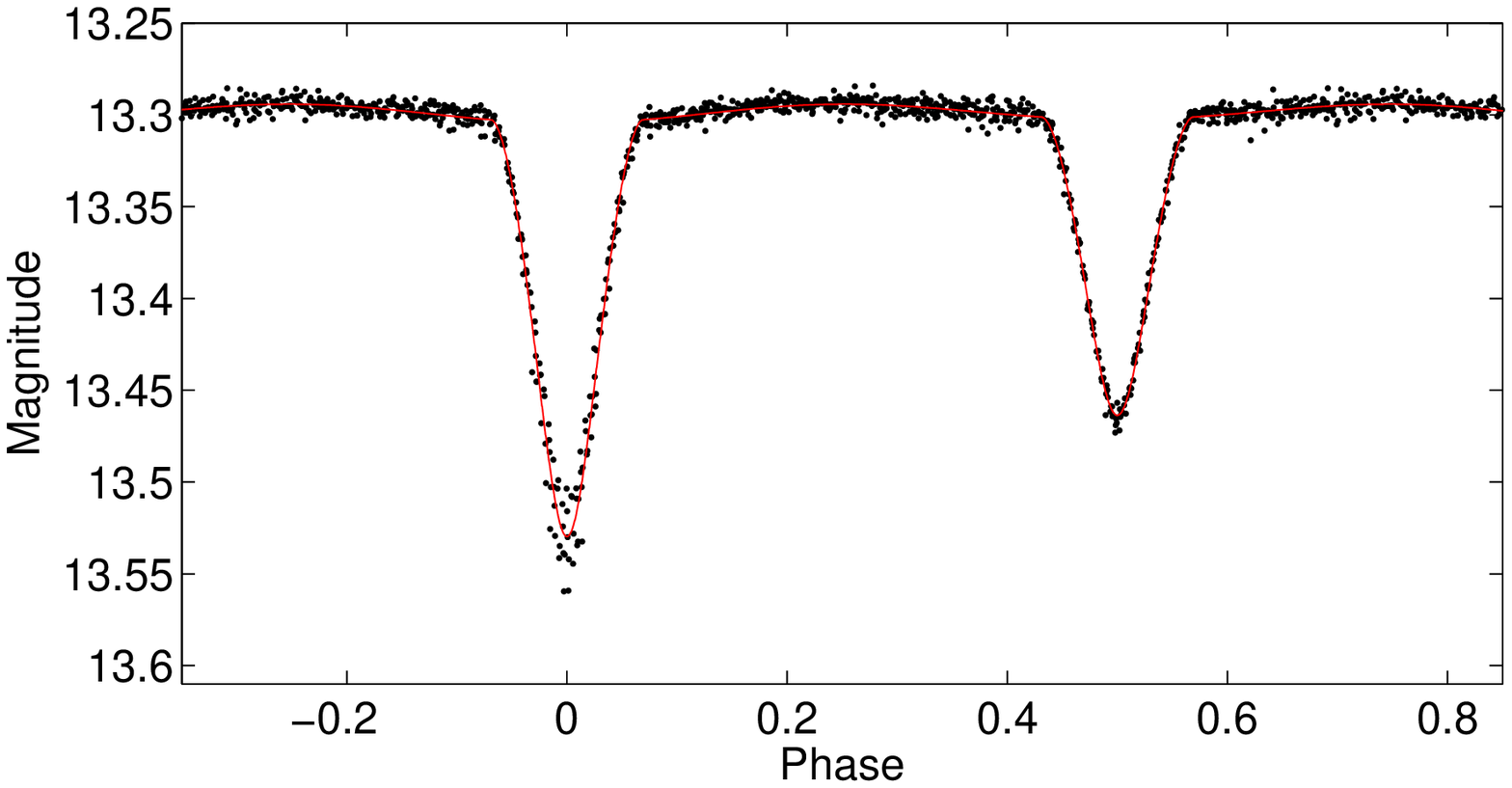}}
 \put(130,80){
 \includegraphics[width=0.24\textwidth]{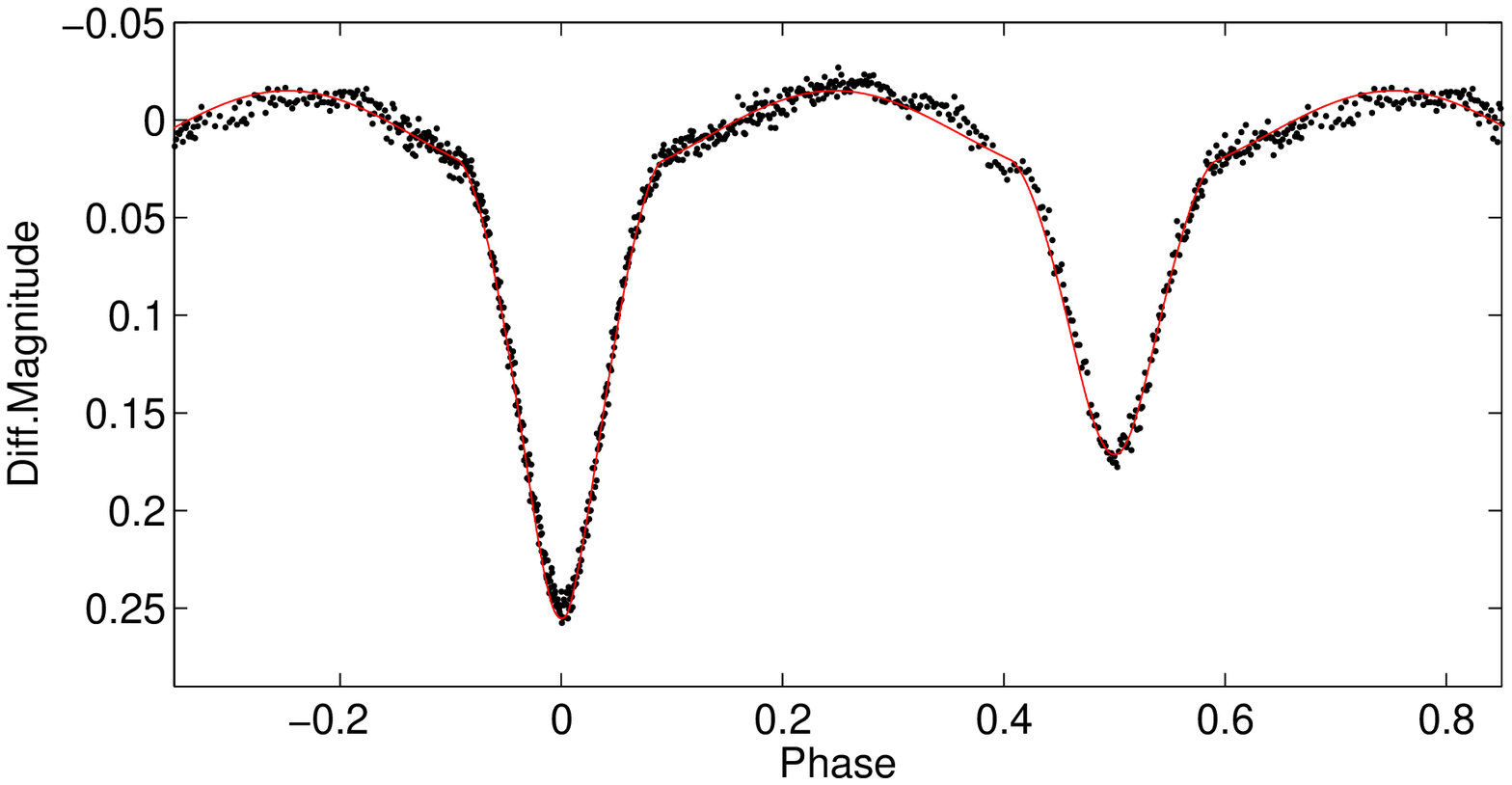}}
  \put(130,0){
 \includegraphics[width=0.24\textwidth]{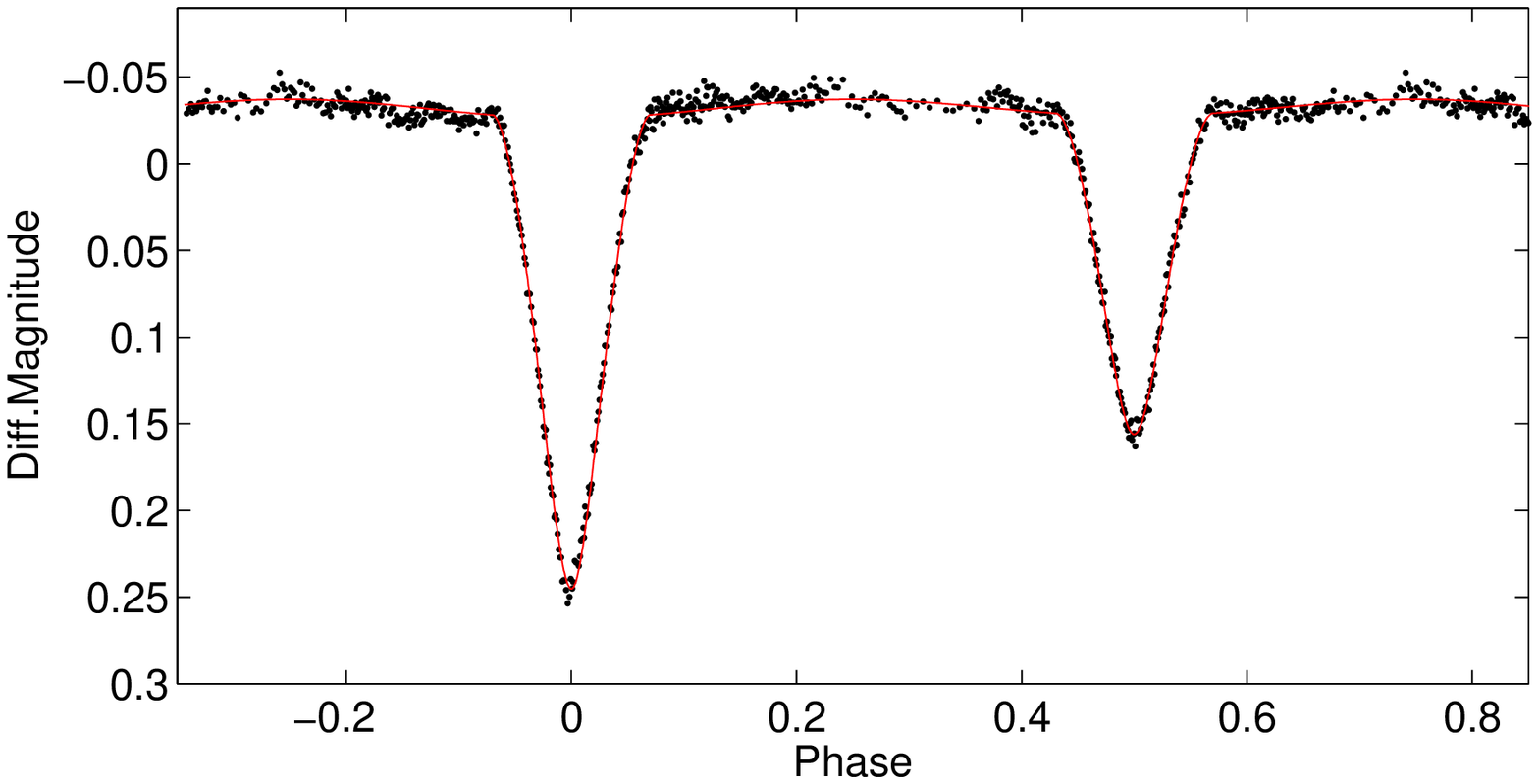}}
  \put(60,95){\small TESS: Pair A}
  \put(60,15){\small TESS: Pair B}
  \put(195,95){\small Z.H.: Pair A}
  \put(195,15){\small Z.H.: Pair B}
 \end{picture}
 \caption{Light-curve fits of CzeV1645 of both eclipsing pairs, as resulting from PHOEBE; see the text for details. TESS data are being compared with our ground-based data obtained by one of the authors (Z.H.).}
 \label{FigLC_czev1645}
\end{figure}

\begin{figure}
 \centering
 \includegraphics[width=0.40\textwidth]{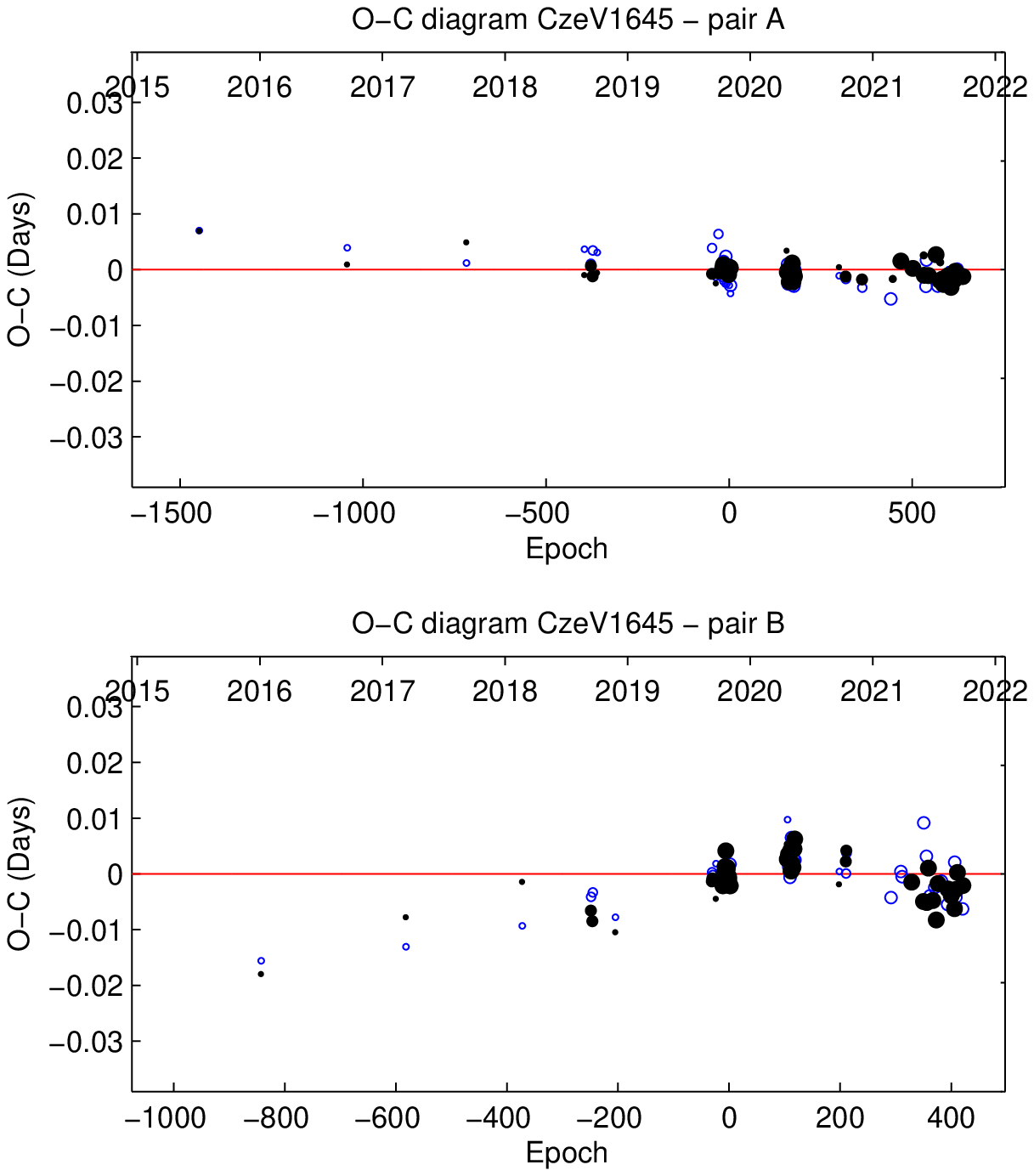}
 \caption{Long-term study of eclipse times of CzeV1645; see the text for details.}
 \label{FigETV_czev1645}
\end{figure}

Firstly, we used the TESS data for our light-curve modelling of both pairs, using a similar approach as
for the previous system. The results with the theoretical fit are given in Fig. \ref{FigLC_czev1645},
while the parameters are written in Table \ref{TabLC}. As one can see, both pairs have
well-covered light curves, and their eclipse depths are also comparable and relatively large. It is
therefore quite surprising that the secondary periodicity was not detected earlier. However, when
downloading and processing the TESS data, we found a problematic issue with that photometry. There were
different trends in the data, which we needed to subtract with a polynomial fitting, but what we were
not able to remove is the different amplitude of pair B in different time epochs. Such an effect is
even more visible when comparing sector 1 and sector 3 of the TESS data (very incomparable eclipse
depths), but an artifact of this is also visible in Fig. \ref{FigLC_czev1645} regarding pair B. What is
quite surprising is the fact that the secondary eclipse of pair B is not affected, but the primary one
is. We have no clear explanation for this strange behavior. Rapid inclination change due to dynamical
effects of the other pair would cause both eclipses to change their depths. An instrumental effect of
the TESS data reduction would be a possible source of this discrepancy. Perhaps a more sophisticated
approach for the TESS data or the emergence of new TESS data is needed here.

For this reason, we also used another source of the photometry. Namely, our new ground-based
data obtained by the author (Z.H.). The comparison of these data with the former TESS ones is being
plotted in Fig. \ref{FigLC_czev1645}. As one can see, both shapes are similar, but not equal (due to
different eclipse depths due to different pixels and apertures used), but their relative precision is
well-comparable. What we found on our ground-based data is some slight asymmetry on the
light curve of pair A (in the out-of-eclipse region close to phases 0.35 and 0.7), possibly explainable
with the presence of a spot on the surface of the star. However, the strange change of the depth of
primary eclipse of pair B is definitely not visible in our data, despite these ranging over a longer
time interval than the TESS ones.

Our light-curve modelling resulted in two well-detached systems of both A and B pairs, but whose two
orbital periods are surprisingly close to a 2:3 mean motion resonance. Therefore, it should be studied
in more detail. Not so surprising is the fact that in using the TESS data, the overall sum of the two
third-light values of both pairs was over 100\%. This is something naturally explainable by the fact of
large TESS pixels, and some additional light from some close-by source, not connected with the
quadruple itself. Nothing like that is visible in our data with a much better angular resolution. There
is some discrepancy between these two solutions (see Table \ref{TabLC}), mainly for pair A. This can
also be caused by larger luminosity contamination of the TESS data, and perhaps also by the fact that
with such a shape of the light curve one cannot reliably derive the mass ratio from photometry alone
due to the small out-of-eclipse variations \citep{2005Ap&SS.296..221T}.

An analysis of all available photometry to derive the eclipse times of both pairs was also carried out.
Besides the TESS data, we also used our new ground-based photometry of the target, together with the
ASAS-SN data and the ZTF as well. On all of these datasets both eclipses are clearly detectable due to
their relatively large depths. This is also the reason why the derived times of eclipses are of good
precision for the ETV analysis (see Fig. \ref{FigETV_czev1645}, showing that practically no visible
variation is evident here). Pair B would show some mild ETV signal, only very barely visible for A.
However, its nature and period is still very hard to estimate. It is probably rather long (over 6
years), and its almost non-detection for pair A would be caused by its lower amplitude (probably due to
the higher mass of pair A). Therefore, we call for another observation of both pairs in the upcoming
years for its final confirmation.

\subsection{CzeV3436}

The only one system in our sample obviously showing two short, periodic, contact-binary-shaped light
curves is CzeV3436 (or also UCAC4 631-070565), which should be visually classified as EW+EW or EB+EW.
However, due to its low brightness we have only very limited information for a proper
analysis. Not only is the effective temperature unsure, but for the distance different
authors also give very different values: 8.2 kpc (GAIA EDR3, \citealt{2021A&A...649A...1G}), 5.7 kpc
\citep{2019AaA...628A..94A}, 5 kpc \citep{2021AJ....161..147B}, and 11.6 kpc
\citep{2019AJ....158..138S}.

\begin{figure}
 \centering
 \begin{picture}(280,150)
 \put(0,80){
 \includegraphics[width=0.24\textwidth]{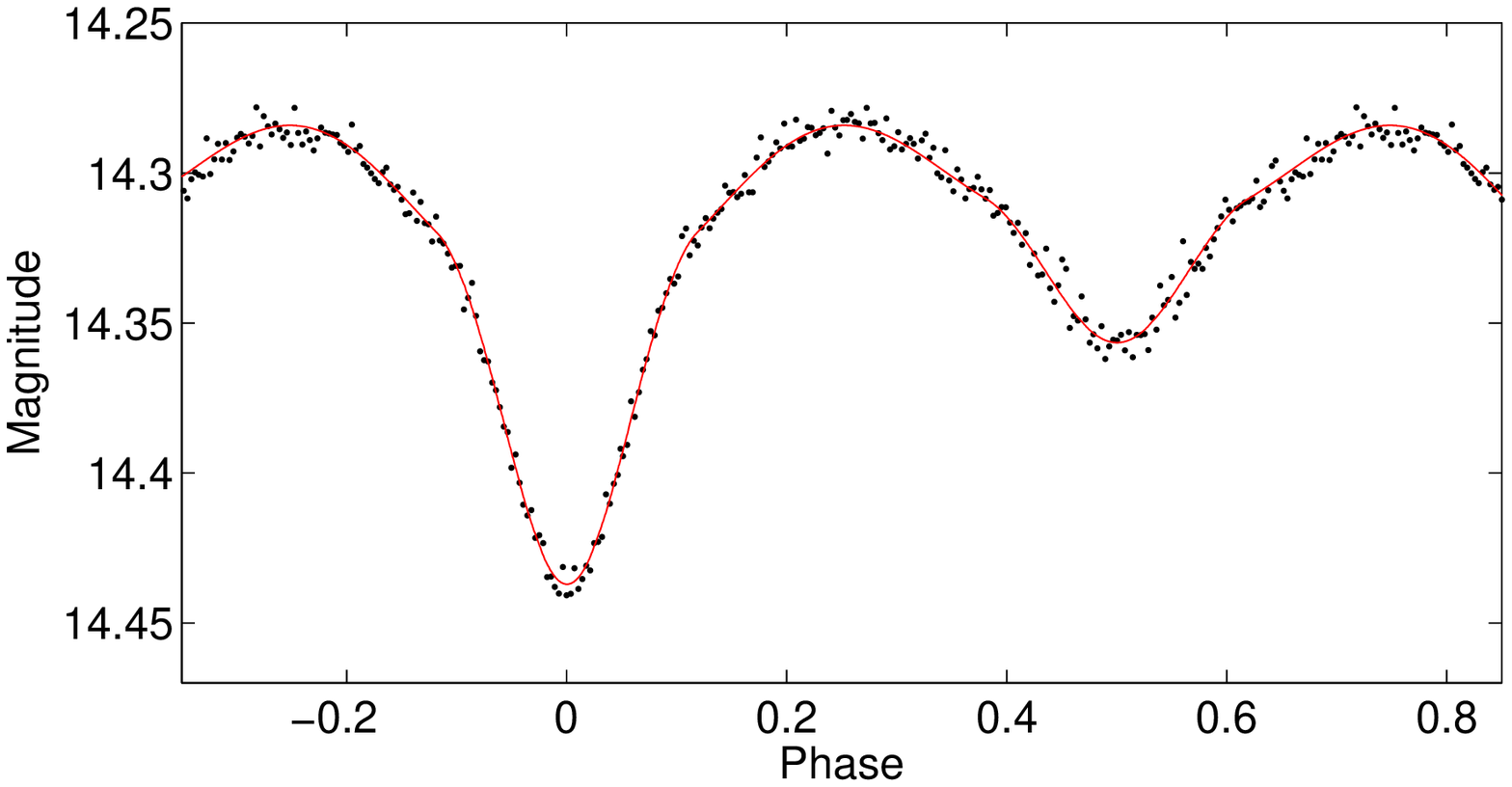}}
 \put(0,0){
 \includegraphics[width=0.24\textwidth]{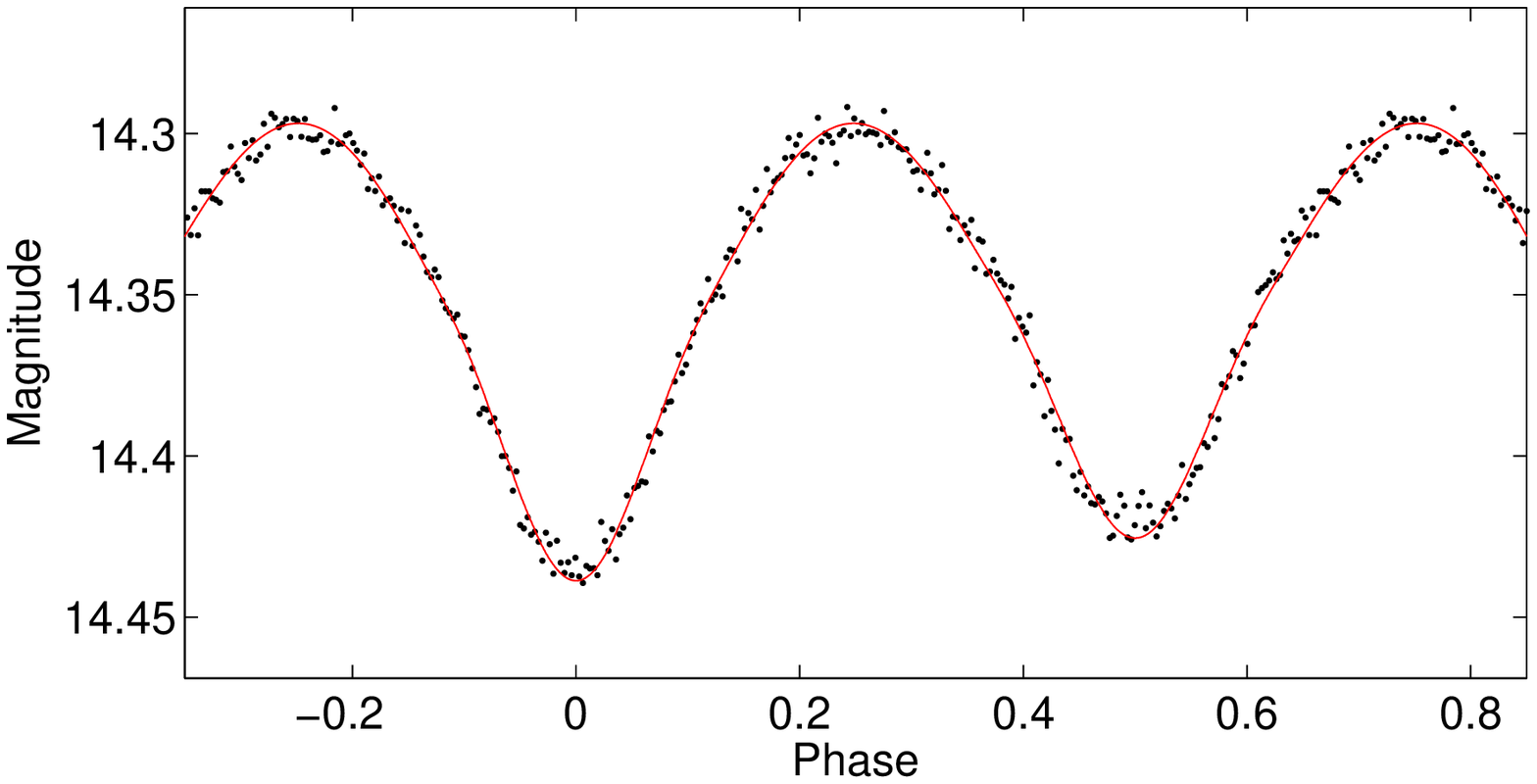}}
 \put(130,80){
 \includegraphics[width=0.24\textwidth]{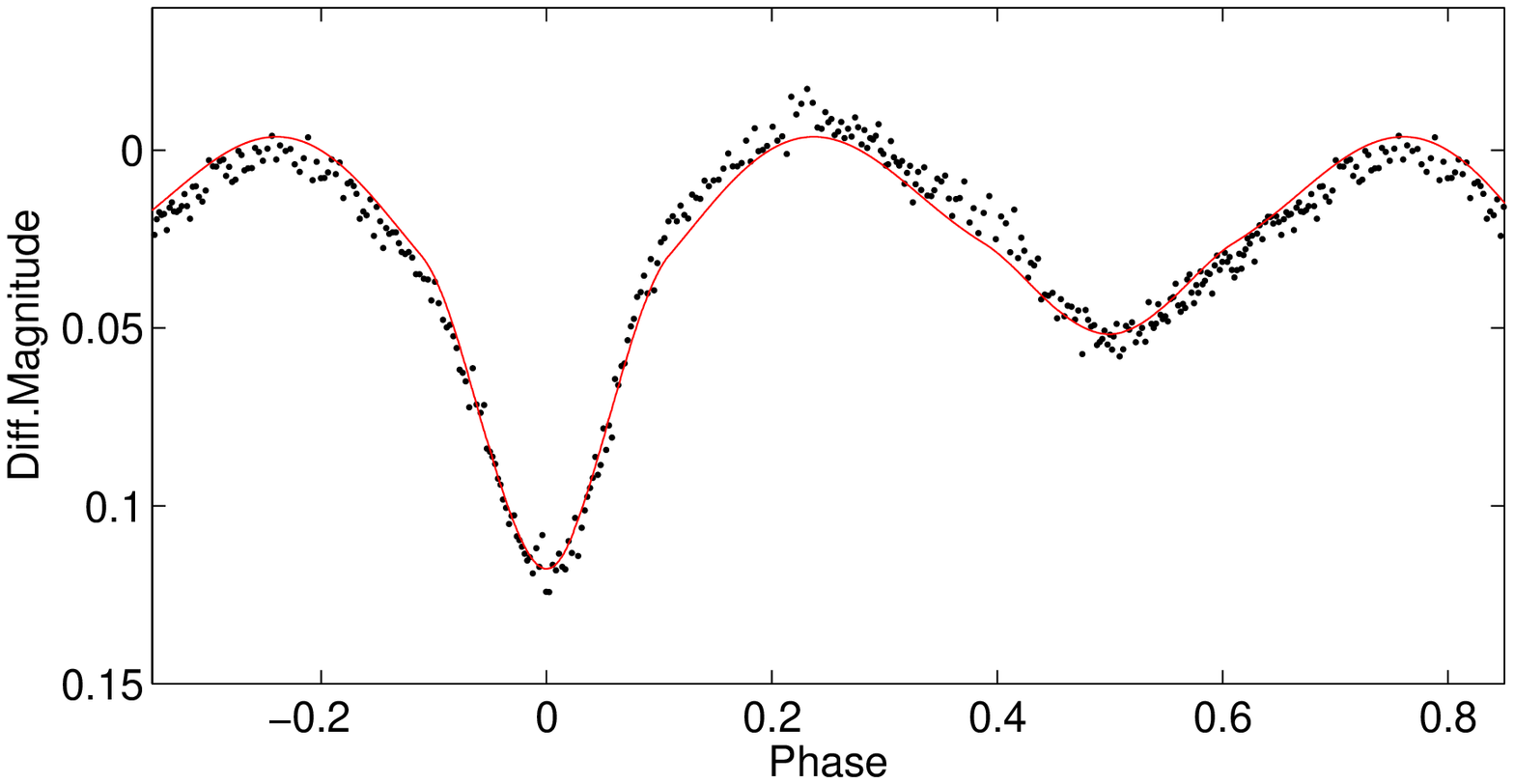}}
  \put(130,0){
 \includegraphics[width=0.24\textwidth]{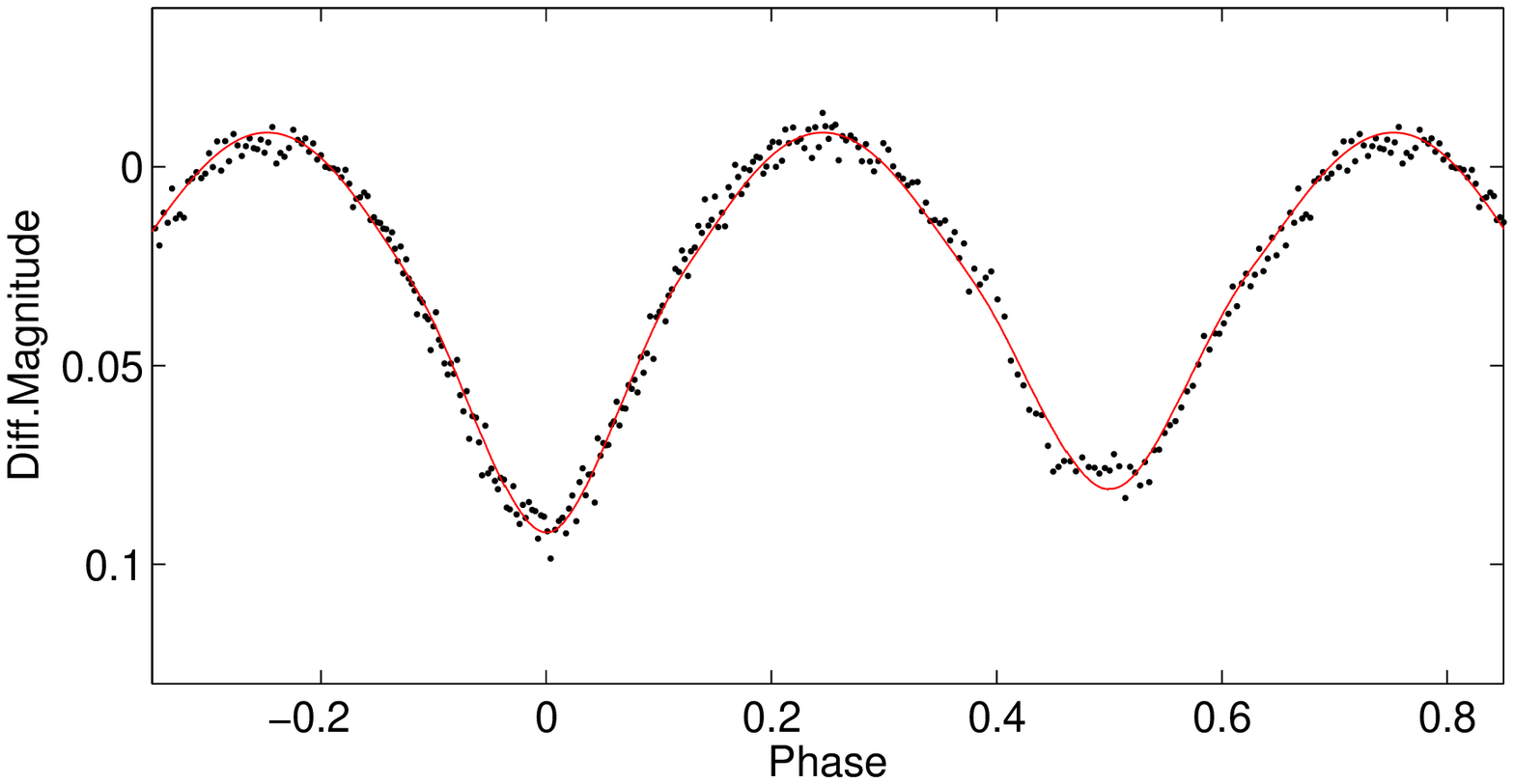}}
  \put(60,95){\small TESS: Pair A}
  \put(60,13){\small TESS: Pair B}
  \put(195,95){\small Z.H.: Pair A}
  \put(195,13){\small Z.H.: Pair B}
 \end{picture}
 \caption{Light-curve fits of CzeV3436 of both eclipsing pairs, as resulted from PHOEBE; see the text for details. TESS data are compared with our ground-based data obtained by one of the authors (Z.H.).}
 \label{FigLC_czev3436}
\end{figure}

There is no variation visible in the ASAS-SN data or on ZTF. This is caused by its faintness
and low amplitude of photometric variations. Hence, the only source of photometric information here is
the TESS satellite and our new ground-based photometry. The analysis was carried out iteratively in
several steps subtracting the particular pair for analysis of the other one. However, after a few
iterations, we finally reached a solution, which we feel is a reasonable one (see Table \ref{TabLC}
for the final parameters and a plot of both curves in Fig. \ref{FigLC_czev3436}). The light curve
solution based on the TESS data provide quite a reasonable picture of the whole system. In agreement
with a similar depth of eclipses of both pairs, their luminosity contributions are also close to 50\%.
However, similarly to the previous case, we found an asymmetric light curve from our ground-based data
here. For this reason, its fit is not as good as the TESS one, and it is affected by larger errors. We
can only speculate about the reason for such deviation.

\begin{figure}
 \centering
 \includegraphics[width=0.40\textwidth]{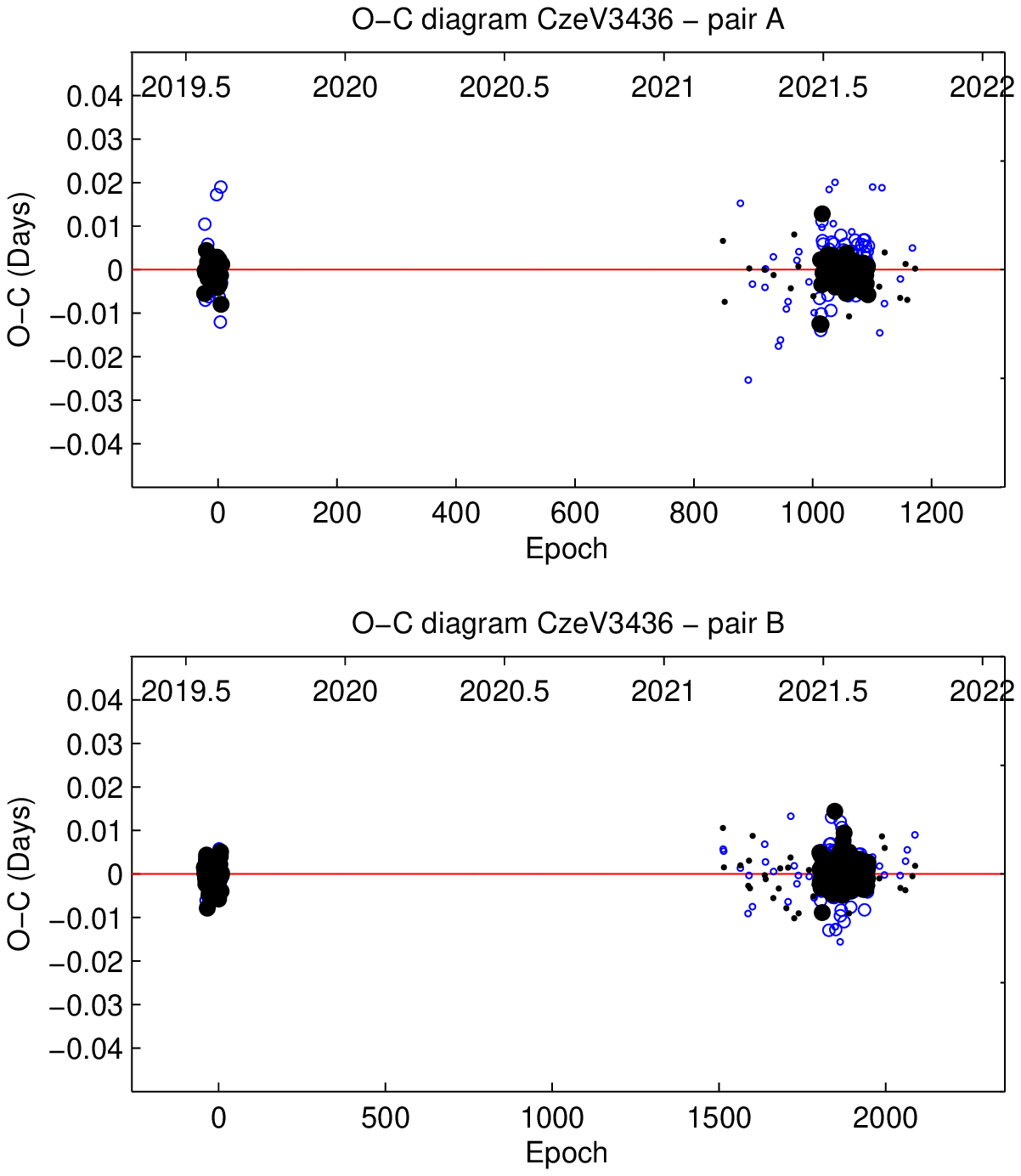}
 \caption{Long-term constancy of period of both pairs of CzeV3436.}
 \label{FigETV_czev3436}
\end{figure}

However, despite all these problems, we also performed the long-term analysis of its period changes.
Both pairs were studied separately, but only based on TESS and our own ground-based data. For this
reason (see Fig. \ref{FigETV_czev3436}), we were not able to trace any visible variation in either of
the pairs. Their mutual orbit should be longer (or face-on oriented orbit), and only future observations
would help to answer the question about their mutual gravitational interaction.

\subsection{OGLE SMC-ECL-1758}

The only system of our Galaxy, located in the SMC field of the OGLE survey, is named OGLE SMC-ECL-1758.
It was also added into our study due to the fact that it was observed several times during our
monitoring of other targets in the SMC fields. Therefore, we have adequately large observational
material besides the OGLE data to perform a thorough analysis of both detected pairs and to reveal its
structure. The star is already known as a doubly eclipsing one since \cite{2013AcA....63..323P}
detected both eclipsing periods. It is quite a typical star in the SMC galaxy, but very little is known
about it at present, and no spectral type estimation was published.

\begin{figure}
 \centering
 \begin{picture}(280,150)
 \put(0,80){
 \includegraphics[width=0.24\textwidth]{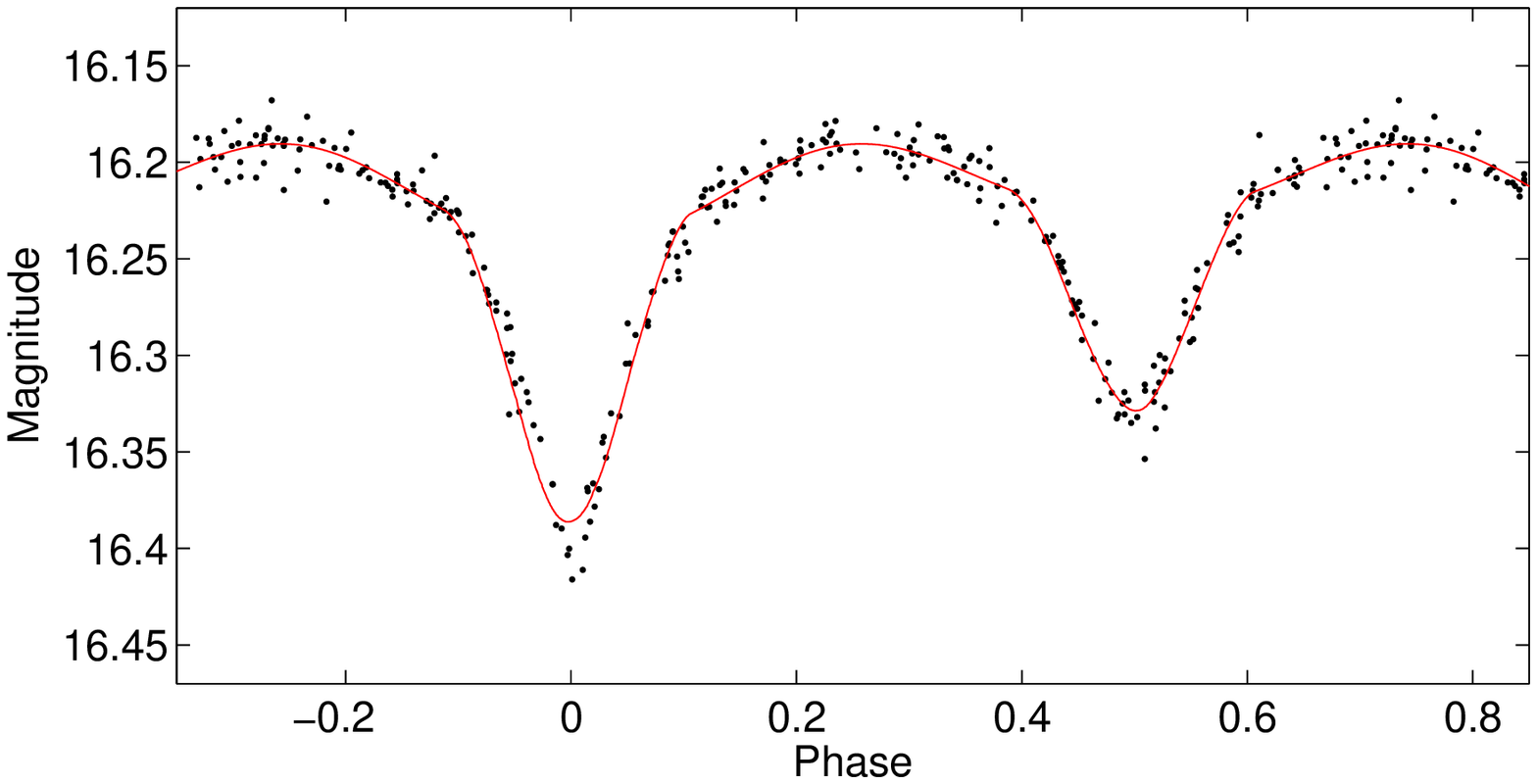}}
 \put(0,0){
 \includegraphics[width=0.24\textwidth]{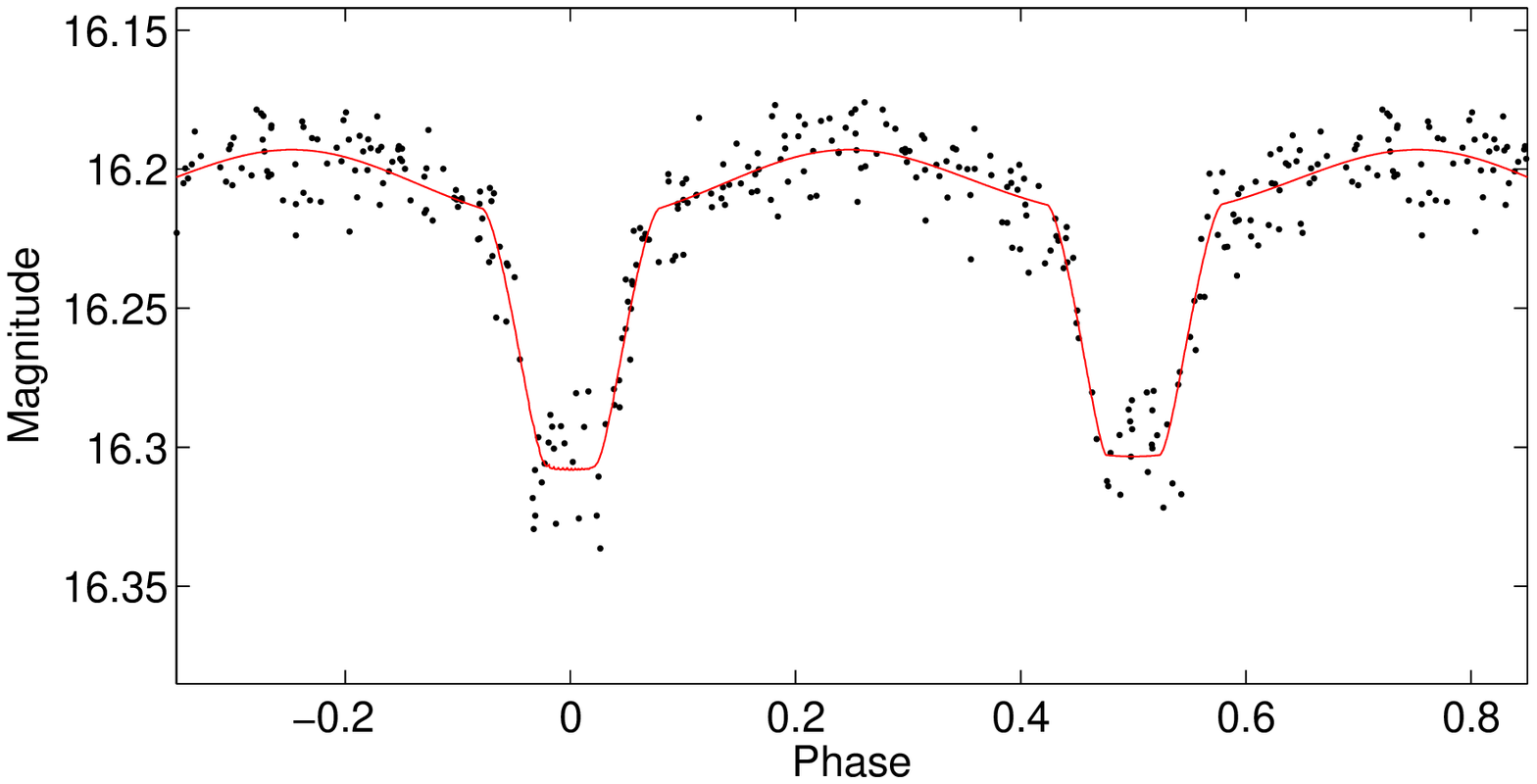}}
 \put(130,80){
 \includegraphics[width=0.24\textwidth]{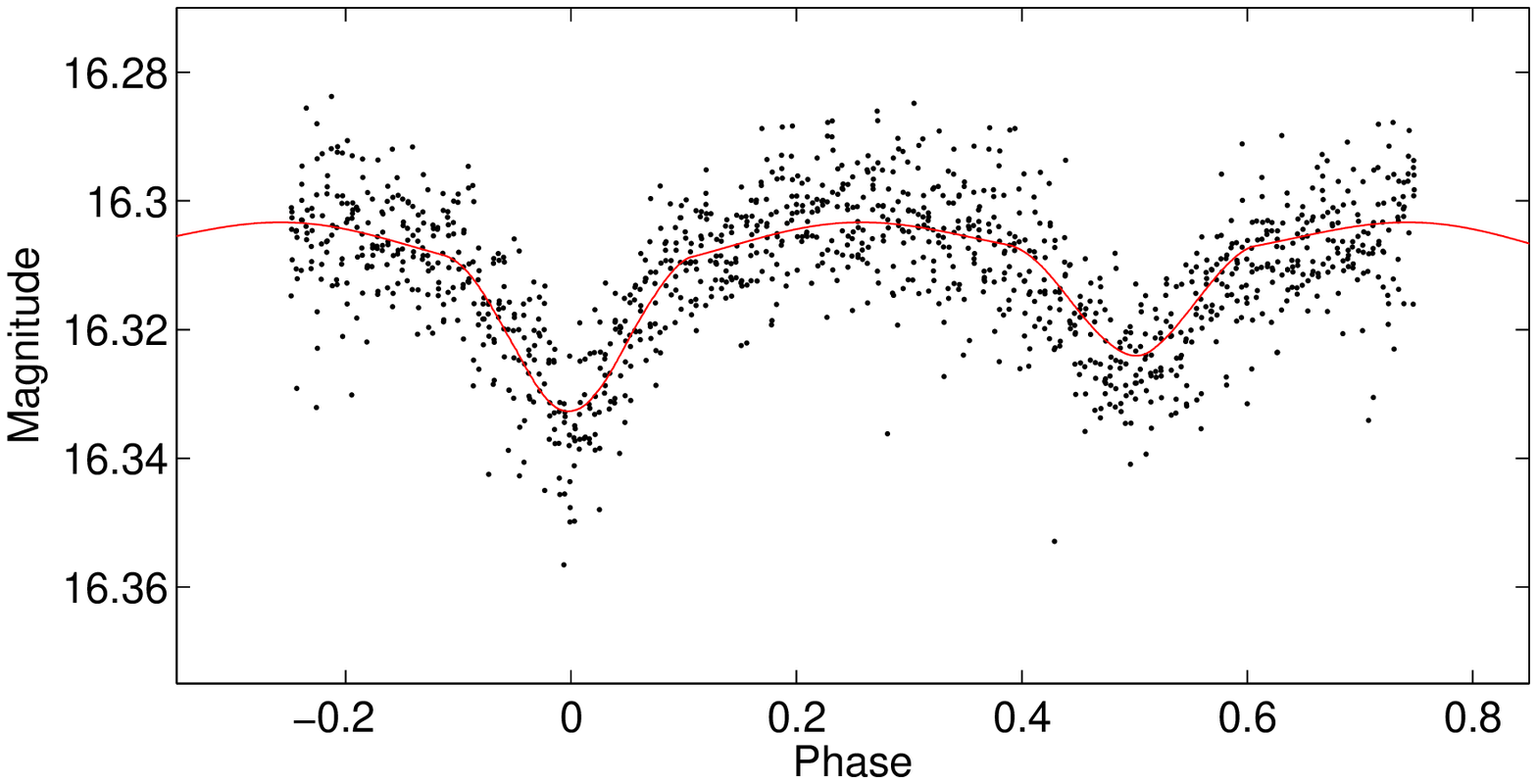}}
  \put(130,0){
 \includegraphics[width=0.24\textwidth]{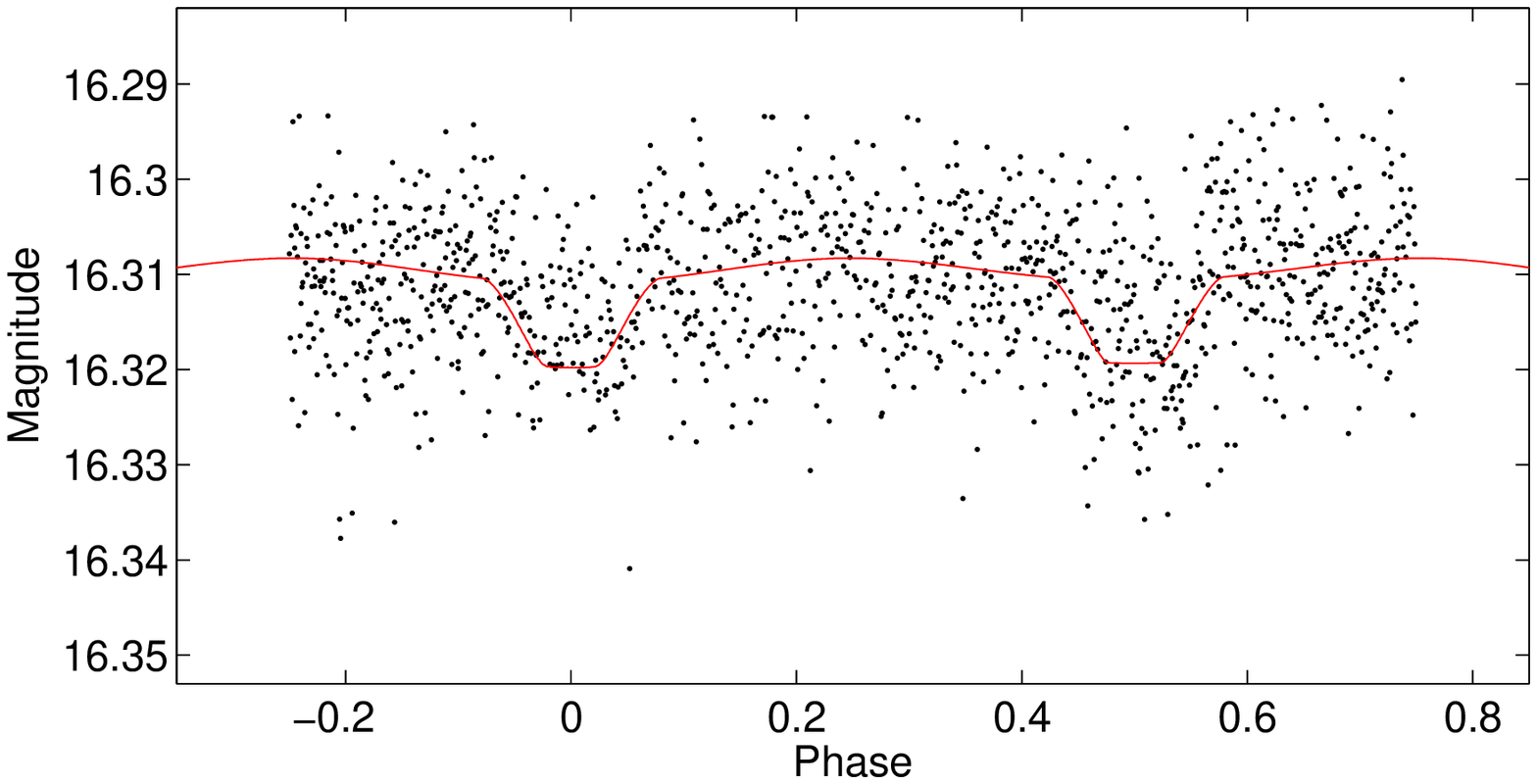}}
  \put(59,95){\small OGLE IV: Pair A}
  \put(59,13){\small OGLE IV: Pair B}
  \put(195,95){\small TESS: Pair A}
  \put(195,13){\small TESS: Pair B}
 \end{picture}
 \caption{Light-curve fits of OGLE SMC-ECL-1758 of both eclipsing pairs, as resulting from PHOEBE; see the text for details. The TESS data are only shown here to compare their quality, and they are not used for the light-curve solution.}
 \label{FigLC_ogle1758}
\end{figure}

We analysed all OGLE data (OGLE II+III+IV phases of the survey), together with the older MACHO data. In
all of these observations, both eclipses are clearly visible because its amplitude is quite large, of
about 0.2 mag (for pair A, slightly less for pair B). The parameters of the light curve fits are given
again in Table \ref{TabLC}, while the fits of the OGLE IV data are plotted in Fig.
\ref{FigLC_ogle1758}. We only used the more precise OGLE IV data for the light-curve modelling due to
their best quality among all available datasets. The observation from MACHO provides more data points,
which are spread over longer time interval, but they are of worse quality. As for the curiosity, we
also plotted the TESS data of the target in Fig. \ref{FigLC_ogle1758} to verify that these data
definitely cannot be used for any light curve modelling. However, we used them at least to derive the
eclipse times of both pairs for construction of the ETV diagrams. The bad quality and uselessness of
the TESS data is due to their low angular resolution and very large pixels. This always causes problems
of contamination of other close-by sources in such dense fields (as this one is in central parts of
SMC), making all variations much smaller. Here the amplitude of photometric variability is about seven
times lower than for the OGLE IV data.

\begin{figure}
 \centering
 \includegraphics[width=0.40\textwidth]{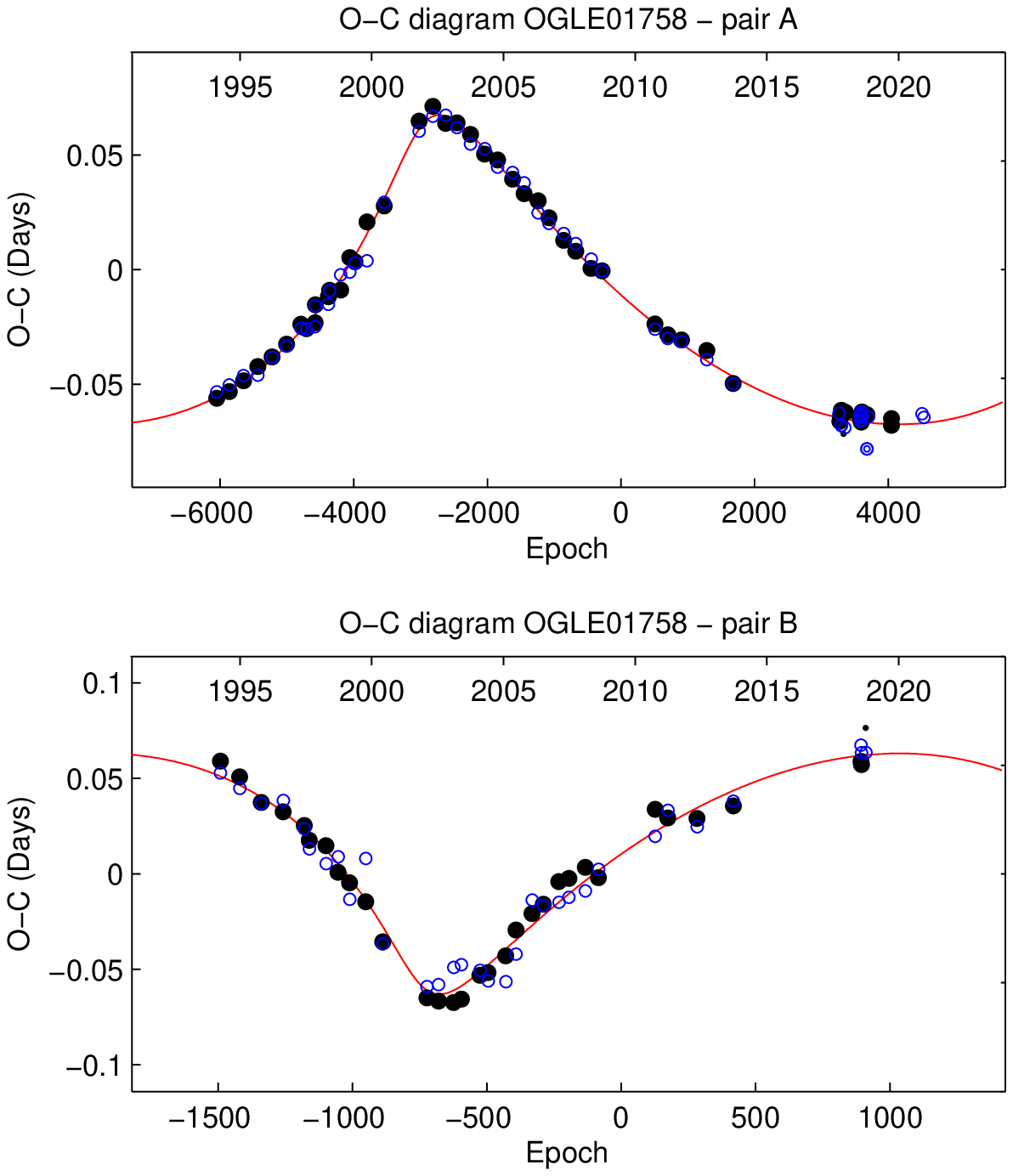}
 \caption{ETV diagram of OGLE SMC-ECL-1758 showing that both inner binaries A and B show significant period variations. This is a consequence of their mutual motion around a common barycenter.}
 \label{Fig_OC_ogle1758}
\end{figure}

Because of much worse quality of the data comparing with the previous three systems, we used a slightly
different approach here. To compute the mass ratio of the eclipsing pairs, the photometry itself is
only slightly sensitive and usually derived with large uncertainty. This especially applies for the
detached systems (see for example the study by \cite{2005Ap&SS.296..221T}). For this reason, we used an
alternative method of mass ratio computation, introduced by \cite{2003MNRAS.342.1334G}. Instead of
directly computing the mass ratio as a free parameter in a light-curve fitting, the mass ratio was
computed from the luminosity ratio of both components (each step of the light-curve fitting in PHOEBE
provides bolometric magnitudes of both components). Naturally, it works with the assumption that both
components are the main sequence stars following a standard mass-luminosity relation. Both of the inner
eclipsing systems were analysed in this way. Quite surprisingly, both of these inner mass ratios were
about the same (of about 0.62--0.63). Moreover, as found from the ETV analysis of both pairs (see Fig.
\ref{Fig_OC_ogle1758}), both pairs are of about the same mass ($q_{A/B} = 0.95$). This is quite a
surprising finding, possibly telling us something about the formation mechanism of such a system. As
stated, for example, in the recent study by \cite{2021Univ....7..352T}, the 2+2 quadruples may be
formed via several mechanisms acting together: core fragmentation, disc instability, dynamical
interactions, and accretion-driven orbital migration. In this aspect, our presented system can shed new
light on these interesting systems and their statistics. Such a system having roughly the same mass of
both binaries (that is $q_{A/B} \approx 1$), but with quite different components within these binaries
($q \approx 0.6$), is not very typical (see for example Fig. 5 in \cite{2021Univ....7..352T}).

Moreover, it was also found that pair B has slightly eccentric orbit (barely detectable at a level
of e$\simeq$0.01, with possible apsidal motion about 100 yr long). However, such a small eccentricity is
barely detectable with the quality of the available data. Finally, it is quite interesting to note that both inner periods, $P_A$, and $P_B,$ are less than 0.5\% away from the 1:4 resonance.
Whether it says something about the resonances themselves, or whether it is pure coincidence, we leave
as an open question.

\section{Discussion and conclusions}  \label{discussion}

We carried out an analysis of four doubly eclipsing systems that had never been studied before, among which
three were discovered and published for the first time here. These new systems are being added
to a still rather limited group of doubly eclipsing systems (today with 159 members). However, it is
interesting that a large majority of these systems are detached and have longer periods, rather than
being short-period contact stars. Two such detached light curves are more easily being disentangled from each
other; hence, their analysis is more straightforward. Moreover, to prove that these are not pulsations
but two real contact binaries can also sometimes be rather tricky. For this reason, we feel that our
system CzeV3436 is quite rare in this aspect. The same also applies for the star CzeV2647, which could
easily be missed as a doubly eclipsing one because of its very low photometric amplitude of pair B,
and only a careful visual inspection of very precise data would be able to detect such a system. All of
these should be taken into account when considering the statistics of these rare systems.

Two out of four analysed systems were detected to orbit around a barycenter, hence constituting a real
2+2 quadruple. The other two systems are waiting for more data to become available in the next few years
to prove such orbit. Due to our limited data coverage in time for all these systems, we believe that
most of these doubly eclipsing systems orbit each other, and only a very limited portion of them
will later be found as chance alignments. For example, for CzeV1645 we believe that new observations in
the upcoming years will confirm our hypothesis of mutual motion and detect the period (maybe
several decades long).

One may also ask whether we can also detect the mutual movement and the orbit using another means. But
spectroscopy needed for such a confirmation would be rather time-consuming due to the detection of
changes in systemic velocities over a longer time interval (because the orbits are rather long), and
the targets are quite faint for good spectroscopy. Another possibility is to detect the two pairs as a
visual double. However, both detected quadruples are rather distant (that is their angular separation
is too small: a few mas for CzeV2647, and below 1 mas for OGLE SMC-ECL-1758), but a more serious
limitation is that both the stars are too faint for good interferometry. Hence, we cannot hope for any
interferometric detection of the two close components to prove our hypothesis independently with
another method. Therefore, the photometry remains the last remaining and most suitable method for
proving such binaries orbit and constitute real quadruples.

\begin{acknowledgements}
An anonymous referee is being acknowledged for useful comments and suggestions, greatly improving the
manuscript. We do thank the NSVS, ZTF, ASAS-SN, TESS, and OGLE teams for making all of the observations
easily public available. The research of P.Z. was supported by the project Progress Q47 {\sc Physics}
of the Charles University in Prague. We are also grateful to the ESO team at the La Silla Observatory
for their help in maintaining and operating the Danish telescope. This work has made use of data from
the European Space Agency (ESA) mission {\it Gaia} (\url{https://www.cosmos.esa.int/gaia}), processed
by the {\it Gaia} Data Processing and Analysis Consortium (DPAC,
\url{https://www.cosmos.esa.int/web/gaia/dpac/consortium}). Funding for the DPAC has been provided by
national institutions, in particular the institutions participating in the {\it Gaia} Multilateral
Agreement. The observations by Z.H. in Velt\v{e}\v{z}e were obtained with a CCD camera kindly borrowed
by the Variable Star and Exoplanet Section of the Czech Astronomical Society. We would like to also
thank M.Ma\v{s}ek and H.Ku\v{c}\'akov\'a for obtaining the photometric data for CzeV2647, and for
observations of OGLE SMC-ECL-1758 J.Merc, J.Jan\'{\i}k, E.Paunzen, and M.Zejda as well. This research
made use of Lightkurve, a Python package for TESS data analysis \citep{2018ascl.soft12013L}. This
research has made use of the SIMBAD and VIZIER databases, operated at CDS, Strasbourg, France and of
NASA Astrophysics Data System Bibliographic Services.
\end{acknowledgements}

\end{document}